\documentclass[fleqn,10pt, aps,pre,nofootinbib, amsmath,amssymb]{wlscirep}

\usepackage[english]{babel}
\usepackage{graphicx}
\usepackage{amsmath,subfigure}
\usepackage{sidecap,comment}
\usepackage{floatrow,color}
\usepackage{tabularx,booktabs}
\usepackage{chemarr}
\usepackage{float,soul}
\usepackage[]{subfigure}
\usepackage{xcolor,colortbl}
\usepackage{nameref,hyperref}

\usepackage{pbox}
\usepackage{refstyle}
\usepackage[symbol]{footmisc}
\usepackage[capitalise]{cleveref}

\usepackage{pdflscape}
\usepackage{afterpage}
\usepackage{capt-of}
\usepackage{rotating}

\title{Characterizing steady states of genome-scale metabolic networks in continuous cell cultures}

\author[1,2]{Jorge Fernandez-de-Cossio-Diaz \thanks{\href{mailto:cossio@cim.sld.cu}{cossio@cim.sld.cu}}}
\author[1]{Kalet León \thanks{\href{mailto:kalet@cim.sld.cu}{kalet@cim.sld.cu}}}
\author[2]{Roberto Mulet \thanks{\href{mailto:mulet@fisica.uh.cu}{mulet@fisica.uh.cu}}}

\affil[1]{Systems Biology Department, Center of Molecular Immunlogy, Havana (Cuba)}
\affil[2]{Group of Complex Systems and Statistical Physics. Department of Theoretical Physics, Physics Faculty, University of Havana (Cuba)}

\begin{abstract}
	In the \emph{continuous} mode of cell culture, a constant flow carrying fresh media replaces culture fluid, cells, nutrients and secreted metabolites. Industrial applications place demands on the steady states attainable in this kind of culture, usually: high-cell density, stability, minimum waste byproduct accumulation, and efficient nutrient use. Here we present a model for continuous cell culture coupling intra-cellular metabolism to extracellular variables describing the state of the bioreactor, taking into account the growth capacity of the cell and the impact of toxic byproduct accumulation. We provide a method to determine the steady states of this system that is tractable for metabolic networks of arbitrary complexity. We demonstrate our approach in a toy model first, and then in a genome-scale metabolic network of the Chinese hamster ovary cell line, obtaining results that are in qualitative agreement with experimental observations. More importantly, we derive a number of consequences from the model that are \emph{independent} of parameter values. First, that the ratio between cell density and dilution rate is an ideal control parameter to fix a steady state with desired metabolic properties invariant across perfusion systems. This conclusion is robust even in the presence of multi-stability, which is explained in our model by the negative feedback loop on cell growth due to toxic byproduct accumulation. Moreover, a complex landscape of steady states in continuous cell culture emerges from our simulations, including multiple metabolic switches, which also explain why cell-line and media benchmarks carried out in batch culture cannot be extrapolated to perfusion. On the other hand, we predict invariance laws between continuous cell cultures with different parameters. A practical consequence is that the chemostat is an ideal experimental model for large-scale high-density perfusion cultures, where the complex landscape of metabolic transitions is faithfully reproduced. Thus, in order to actually reflect the expected behavior in perfusion, performance benchmarks of cell-lines and culture media should be carried out in a chemostat.
\end{abstract}

\begin{document}

\flushbottom
\maketitle

\section{Introduction}

Biotechnological products are obtained by treating cells as little factories that transform substrates into products of interest. There are three major modes of cell culture: batch, fed-batch and continuous. In batch, cells are grown with a fixed initial pool of nutrients until they starve, while in fed-batch the pool of nutrients is re-supplied at discrete time intervals. Cell cultures in the continuous mode are carried out with a constant flow carrying fresh medium replacing culture fluid, cells, unused nutrients and secreted metabolites, usually maintaining a constant culture volume. While at present most biotechnology industrial facilities adopt batch or fed-batch processes, the advantages of continuous processing have been vigorously defended in the literature \cite{Werner1992, Griffiths1992a, Kadouri1997, Werner1998, Croughan2015}, and currently some predict its widespread adoption in the near future \cite{Konstantinov2015}.

A classical example of continuous cell culture is the chemostat, invented in 1950 independently by Aaron Novick and Leo Szilard \cite{Novick1950} (who also coined the term \emph{chemostat}) and by Jacques Monod \cite{Monod1950}. In this system, microorganisms reside inside a vessel of constant volume, while sterile media, containing nutrients essential for cell growth, is delivered at a constant rate. Culture medium containing cells, remanent substrates and products secreted by the cells are removed at the same rate, maintaining a constant culture volume. The main dynamical variable in this system is the \emph{dilution rate} ($D$), which is the rate at which culture fluid is replaced divided by the culture volume. In a well-stirred tank any entity (molecule or cell) has a probability per unit time $D$ of leaving the vessel. In industrial settings, higher cell densities are achieved by attaching a cell retention device to the chemostat, but allowing a bleeding rate to remove cell debris \cite{Castilho2002}. Effectively only a fraction $0 \le \phi \le 1$ of cells are carried away by the output flow $D$. This variation of the continuous mode is known as \emph{perfusion culture}.

By definition, a continuous cell culture ideally reaches a steady state when the macroscopic properties of the tank (cell density and metabolite concentrations) attain stationary values. Industrial applications place demands on the steady state, usually: high-cell density, minimum waste byproduct accumulation, and efficient nutrient use. However, identical external conditions (dilution rate, media formulation) may lead to distinct steady states with different metabolic properties (a phenomenon known in the literature as \emph{multi-stability} or \emph{multiplicity of steady states}) \cite{Europa2000, Follstad1999, Altamirano2001, Hayter1992, Gambhir2003}. Therefore, for the industry, it becomes fundamental to know in advance, given the cell of interest and the substrates to be used, which are the possible steady states of the system and how to reach them. Moreover, to satisfy production demands, it may be advantageous to extend the duration of a desired steady state indefinitely \cite{Konstantinov2015}, implying that their stability properties are also of great interest.

Fortunately, in the last few years it has been possible to exploit an increasingly available amount of information about cellular metabolism at the stoichiometric level to build genome-scale metabolic networks  \cite{Kanehisa2014,Caspi2014}.
These networks have been modeled by different approaches \cite{Noor2016a,Lewis2012} but Flux Balance Analysis (FBA) has been particularly successful predicting cell metabolism in the growth phase \cite{Ibarra2002}. FBA starts assuming a quasi-steady state of intra-cellular metabolite concentrations, which is easily translated into a linear system of balance equations to be satisfied by reaction fluxes. This system of equations is under-determined and a biologically motivated metabolic objective, such as biomass synthesis, is usually optimized to determine the complete distribution of fluxes through the solution of a Linear Programming problem \cite{Palsson2006}. This approach was first used to characterize the metabolism of bacterial growth \cite{Edwards2001}, but  later has  been applied also to eukaryotic cells \cite{Shlomi2011, Jouhten2012}. Alternatively, given a set of under-determined linear equations, one can estimate the space of feasible solutions of the system and average values of the reaction fluxes \cite{Braunstein2017, Braunstein2008, Fernandez-de-Cossio-Diaz2016}.

To consider the temporal evolution of a culture, FBA may be applied to successive points in time, coupling cell metabolism to the dynamics of extra-cellular concentrations. This is the approach of \emph{Dynamic Flux Balance Analysis} (DFBA) \cite{Mahadevan2002} and has been applied prominently either to the modeling of batch/fed-batch cultures or to transient responses in continuous cultures, being particularly successful in predicting metabolic transitions in E. Coli and yeast \cite{Mahadevan2002, Meadows2010,Jouhten2012}. However, to the best of our knowledge, the steady states of  continuous cell cultures have not been investigated before. First, because DBFA for genome-scale metabolic networks may be a computational demanding task, particularly when the interest is to understand long-time behavior. Second, because it assumes knowledge of kinetic parameters describing metabolic exchanges between the cell and culture medium, that are usually unknown in realistic networks. Moreover, although the importance of toxic byproduct accumulation has been appreciated for decades \cite{Visek1968, Hassell1991}, its impact on steady states of continuous cultures has been studied mostly in simple metabolic models involving few substrates \cite{Guthke1980, Hegewald1981}, while it has been completely overlooked in DFBA of large metabolic networks. Lactate and ammonia are the most notable examples in this regard and have been widely studied in experiments in batch and continuous cultures \cite{Ozturk1992, Bakker1996, Hu1997, Schneider1996, Hassell1991}.

Our goal in this work is to introduce a detailed characterization of the steady states of cell cultures in continuous mode, considering the impact of toxic byproduct accumulation on the culture, and employing a minimum number of essential kinetic parameters. To achieve this and inspired by the success of DFBA in other settings we couple macroscopic variables of the bioreactor (metabolite concentrations, cell density) to intracellular metabolism. However, we explain how to proceed directly to the determination of steady states, bypassing the necessity of solving the dynamical equations of the problem. This spares us from long simulation times and provides an informative overview of the dynamic landscape of the system. The approach, presented here for a toy model and for a genome-scale metabolic network of CHO-K1, but easily extensible to other systems, supports the idea that multi-stability, \emph{i.e.}, the coexistence of multiple steady states under identical external conditions, arises as a consequence of toxic byproduct accumulation in the culture. We find and characterize specific transitions, defined by simultaneous changes in the effective cell growth rate and metabolic states of the cell, and find a wide qualitative agreement with experimental results in the literature. Our analysis implies that batch cultures, typically used as benchmarks of cell-lines and culture media, are unable to characterize the landscape of metabolic transitions exhibited by perfusion systems. On the other hand, our results suggest a general scaling law that translates between the steady states of a chemostat and any perfusion system. Therefore, we predict that the chemostat is an ideal experimental model of high-cell density perfusion cultures, enabling a faithful characterization of the performance of a cell-line and media formulation truly valid in perfusion systems.

\section{Materials and methods}

\subsection{Dynamical model of the perfusion system}

\begin{figure}
	\centering
	\includegraphics[width=9cm]{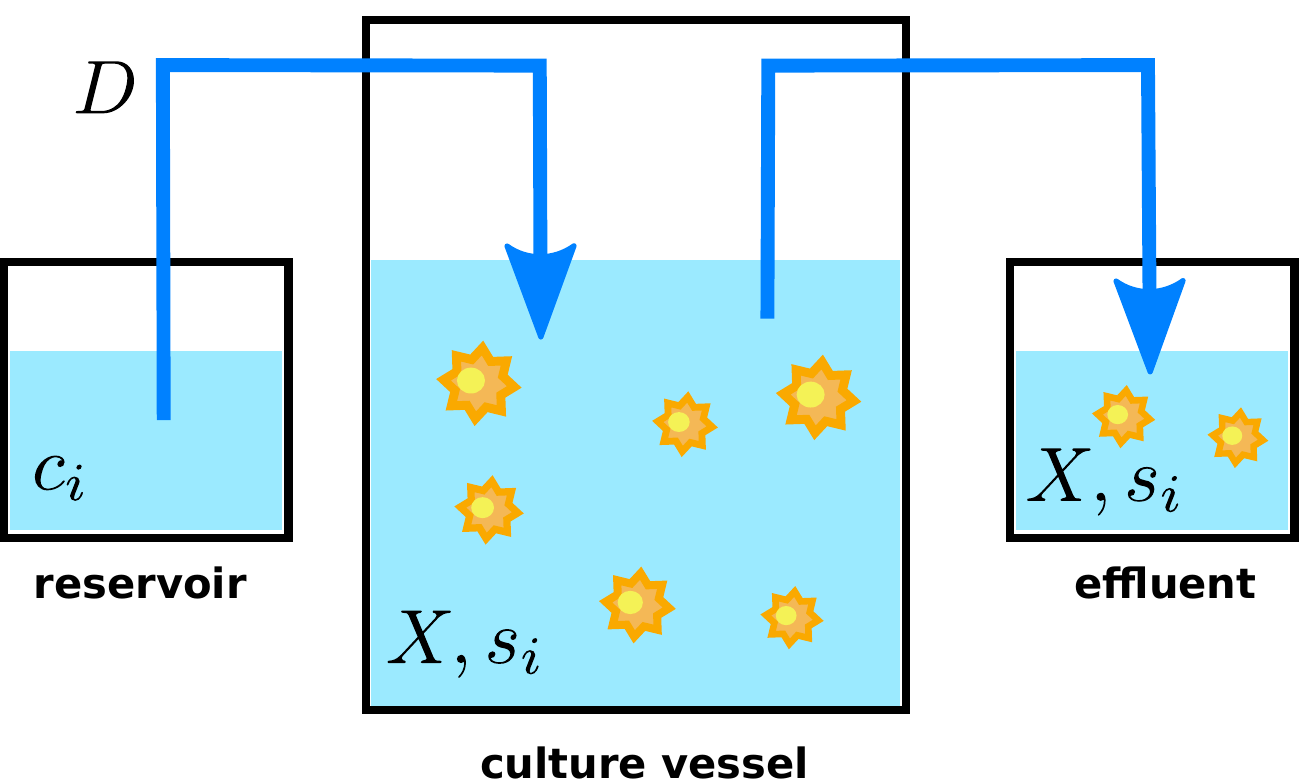}
	\caption{{\bf Cell culture in continuous mode.} A cell culture is grown in a vessel that is continuously fed with a constant flux (blue arrows) of fresh media coming from a reservoir. An equivalent flux carries used media and cells away from the culture vessel, maintaining a constant volume in the culture. The effluent contains cells, secreted metabolites and unused substrates. The figure displays the simple case of no cell retention, where the cell density in the effluent is the same as in the culture. Notation: substrate concentrations in media reservoir ($c_i$), cell density and metabolite concentrations in the culture ($X,s_i$), dilution rate ($D = F / \nu$, where $F$ is the input/output flux and $\nu$ the culture volume).}
\label{fig:tank}
\end{figure}

We study an homogeneous population of cells growing inside a well-mixed bioreactor \cite{BenYahia2015}, where fresh medium continuously replaces culture fluid (\cref{fig:tank}). The fundamental dynamical equations describing this system are:
\begin{align}
\frac{\mathrm{d} X}{\mathrm{d}t} &= (\mu - \phi D) X 
\label{eq:dX} \\
\frac{\mathrm{d} s_i}{\mathrm{d}t} &= -u_i X - (s_i - c_i)D
\label{eq:ds}
\end{align}
where $X$ denotes the density of cells in the bioreactor (units: gDW / L), $\mu$ the effective cell growth rate (units: 1 / hr), $u_i$ the specific uptake of metabolite $i$ (units: mmol/gDW/hr), and $s_i$ the concentration of metabolite $i$ in the culture (units: mM). The external parameters controlling the culture are the medium concentration of metabolite $i$, $c_i$ (units: mM), the dilution rate, $D$ (units: 1 / day), and the bleeding coefficient, $\phi$ (unitless), which in perfusion systems characterizes the fraction of cells that escape from the culture through a cell-retention device \cite{Castilho2002} or a bleeding rate. For convenience of notation, in what follows an underlined symbol like $\underline s$ will denote the vector with components $\{s_i\}$.

Equation \ref{eq:dX} describes the dynamics of the cell density as a balance between cell growth and dilution, while \cref{eq:ds} describes the dynamics of metabolite concentrations in the culture as a balance between cell consumption (or excretion if $u_i < 0$) and dilution. One must notice that at variance with the standard formulation of DFBA, the last terms in the right-hand side of both equations enable the existence of non-trivial steady states (with non-zero cell density) which are impossible in batch. These are the steady states that are relevant for industrial applications adopting the perfusion model and that we study in what follows.

Still, we require a functional connection between variables describing the macroscopic state of the tank ($X$, $\underline s$) and the average behavior of cells ($\underline u$, $\mu$). We start assuming that metabolites inside the cell attain quasi-steady state concentrations \cite{Edwards2002}, so that fluxes of intra-cellular metabolic reactions balance at each metabolite. If $N_{ik}$ denotes the stoichiometric coefficient of metabolite $i$ in reaction $k$ ($N_{ik} > 0$ if metabolite $i$ is produced in the reaction, $N_{ik}<0$ if it is consumed), and $r_k$ is the flux of reaction $k$, then the metabolic network produces a net output flux of metabolite $i$ at a rate $\sum_k N_{ik}r_k$, where $N_{ik} = 0$ if metabolite $i$ does not participate in reaction $k$. This output flux must balance the cellular demands for metabolite $i$. In particular we consider a constant maintenance demand at rate $e_i$ which is independent of growth \cite{Kilburn1969,Sheikh2005}, as well as the requirements of each metabolite for the synthesis of biomass components. If $y_i$ units of metabolite $i$ are needed per unit of biomass produced \cite{Feist2010, Feist2016}, and biomass is synthesized at a rate $z$, we obtain the following overall balance equation for each metabolite $i$:
\begin{equation}
\sum_k N_{ik} r_k + u_i = e_i + y_i z,\quad \forall i
\label{eq:balance}
\end{equation}
It is also well known that a cell has a limited enzymatic budget \cite{Noor2016a}. The synthesis of new enzymes, needed to catalyze many intracellular reactions, consumes limited resources, including amino acids, energy, cytosolic \cite{Beg2007b, Shlomi2011, Vazquez2016} or membrane space \cite{Zhuang2011} (for enzymes located on membranes), ribosomes \cite{Basan2015b}, all of which can be modeled as generic enzyme costs \cite{Noor2016a}. We split reversible reaction fluxes into negative and positive parts, $r_k = r_k^+ - r_k^-$, with $r_k^\pm \ge 0$, and quantify the total cost of a flux distribution in the simplest (approximate) linear form \cite{Noor2016a}:
\begin{equation}
\alpha = \sum_k (\alpha_k^+ r_k^+ + \alpha_k^- r_k^-) \le C
\label{eq:crowd}
\end{equation}
where $\alpha_k^+, \alpha_k^-$ are constant flux costs. The limited budget of the cell to support enzymatic reactions is modeled as a constrain $\alpha \le C$, where $C$ is a constant maximum cost. Thermodynamics places additional reversibility constrains on the flux directions of some intra-cellular reactions \cite{Beard2008}, which can be written as:
\begin{equation}
\mathrm{lb}_k \le r_k \le \mathrm{ub}_k,\quad \mathrm{lb}_k,\mathrm{ub}_k \in \{-\infty, 0, \infty\}.
\label{eq:lbub}
\end{equation}
Additionally, some uptakes $u_i$ are limited by the availability of nutrients in the culture. We distinguish two regimes. If the cell density is low, nutrients will be in excess and uptakes are only bounded by the intrinsic kinetics of cellular transporters. In this case $u_i \le V_i$, where $V_i$ is a constant maximum uptake rate determined by molecular details of the transport process. These will be the only kinetic parameters introduced in the model. When the cell density increases and the concentrations of limiting substrates reach very low levels, a new regime appears where cells compete for resources. In this regime the natural condition $s_i \ge 0$ together with the mass balance equation (\cref{eq:ds} in steady state) imply that $u_i \le c_i D/X$.
In summary,
\begin{equation}
-L_i \le u_i \le \min\{V_i, c_i D/X\}
\label{eq:u}
\end{equation}
where $L_i = 0$ for metabolites that cannot be secreted, and $L_i = \infty$ otherwise. Thus, an important approximation in our model is that low concentrations of limiting nutrients are replaced by an exact zero. The ratio $D/X$ in \cref{eq:u} establishes the desired coupling between internal metabolism and external variables of the bioreactor.

Next, we reason that, although cellular clones in biotechnology are artificially chosen according to various productivity-related criteria \cite{Wurm2004}, the growth rate is typically under an implicit selective pressure. We will consider then that the  flux distribution of metabolic reactions inside the cell maximizes the rate of biomass synthesis, $z$, subject to all the constrains enumerated above. Note that to carry out this optimization it is enough to solve a linear programming problem, for which efficient algorithms are available \cite{Vanderbei2014}. This formulation is closely related to Flux Balance Analysis (FBA) \cite{Edwards2001, Palsson2006, Varma1994a, Orth2010a}, but some of the constrains imposed here might be unfamiliar. In particular, \cref{eq:crowd} has been used before to explain switches between high-yield and high-rate metabolic modes under the name \emph{FBA with molecular crowding} (FBAwMC) \cite{Beg2007b, Vazquez2010, Meiser2016b}, while the right-hand side of \cref{eq:u} is a novel constrain introduced in this work to model continuous cell cultures. If multiple metabolic flux distributions are consistent with a maximal biomass synthesis rate \cite{Mahadevan2003}, the one with minimum cost $\alpha$ (cf. \cref{eq:crowd}) is selected \cite{Noor2016a}. Summarizing, from the complete solution of the linear program we obtain the optimal $z$, and the metabolic fluxes $\underline u$ feeding the synthesis of biomass.

Finally, the net growth rate of cells $\mu$ (see \cref{eq:dX}) is essentially determined by the cellular capacity to synthesize biomass (rate $z$), but it may also be affected by environmental toxicity. In the examples presented below we considered that: $\mu = z - \sigma(\underline s)$ or $\mu = z \times K(\underline s)$, corresponding to two different mechanisms explored in the literature \cite{Schneider1996, Martinelle1993}. In the first case $\sigma(\underline s)$ is easily interpreted as the death rate of the cell, while $0 \le 1 - K(\underline s) \le 1$ represents a fraction of biomass that must be expended on non-growth related activities, for example, due to increased maintenance demands on account of environmental toxicity (but see also Refs.\cite{Bertolazzi2005, Portner1996} and in particular B. Ben Yahia \emph{et al.} \cite{BenYahia2015} for a recent review of the subject). Both $\sigma(\underline s)$ and $K(\underline s)$ depend on the concentrations of toxic metabolites in the culture, such as lactate and ammonia.

\subsection{Metabolic networks}

\paragraph{Toy model}

\begin{figure}
	\centering
	\includegraphics[width=5cm]{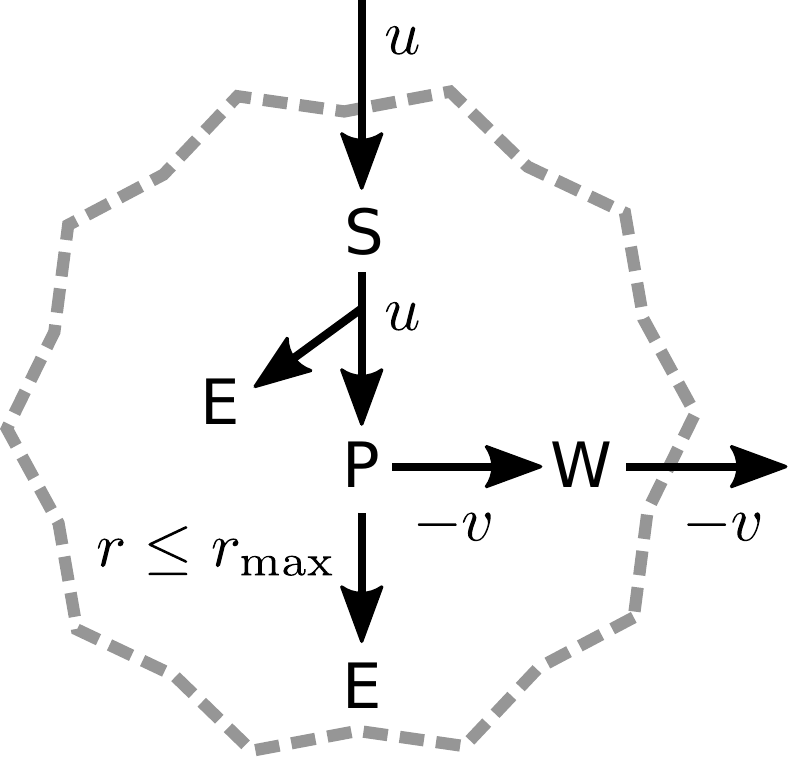}
	\caption{{\bf Diagram of a simple metabolic network.} A primary carbon source $\mathrm{S}$ is consumed by the cell at a rate $u \ge 0$. It is processed into an intermediate $\mathrm{P}$, generating $N_\mathrm{F}$ energy units per unit $\mathrm{S}$ consumed. The intermediate $\mathrm{P}$ can be excreted in the form of a waste product $W$ (rate $-v \le 0$), or it can be completely oxidized (rate $r\ge0$) generating an additional $N_\mathrm{R}$ energy units ($N_\mathrm{R} \gg N_\mathrm{F}$). The respiration rate is capped, $r \le r_\mathrm{max}$.} 
	\label{fig:vazquez-net}
\end{figure}

To gain insight into the kind of solutions expected, we examined first a simple metabolic network that admits an analytic solution. It is based on the simplified network studied by A. Vazquez \emph{et al.} to explain the Warburg effect \cite{Vazquez2010b}, and serves as a minimal model of metabolic transitions in the cell \cite{Capuani2015,aberrant,Vazquez2010b}.

A diagram of the network is shown in \cref{fig:vazquez-net}. There are four metabolites: a primary nutrient $\mathrm{S}$, an energetic currency $\mathrm{E}$, an intermediate $\mathrm{P}$, and a waste product $\mathrm{W}$. Only $\mathrm S$ and $\mathrm W$ can be exchanged with the extracellular medium, and their concentrations in the tank will be denoted by $s$ and $w$, respectively. The cell can consume $\mathrm S$ from the medium at a rate $u \ge 0$. The nutrient is first processed into $\mathrm{P}$, generating $N_\mathrm F$ units of $\mathrm E$ per unit $\mathrm S$ processed. The intermediate can have two destinies: it can be excreted in the form of $\mathrm{W}$ (rate $-v \le 0$), or it can be further oxidized (rate $r \ge 0$), generating $N_\mathrm{R}$ units of $\mathrm{E}$ per unit $\mathrm{P}$. These two pathways are reminiscent of fermentation and respiration. We assume that $N_\mathrm{F} \ll N_\mathrm{R}$, which is consistent with the universally lower energy yield of fermentation versus respiration. Therefore, a maximization of energy output implies that the respiration mode is preferred. However, the enzymatic costs required to enable respiration are very high compared to fermentation. Therefore in \cref{eq:crowd} only the costs of respiration are significant \cite{Vazquez2010b}, which implies that this flux is bounded:
\begin{equation}
r \le r_\mathrm{max}
\label{eq:rmax}
\end{equation}
Metabolic overflow occurs when the nutrient uptake is higher than the respiratory bound $r_\mathrm{max}$. The remaining $\mathrm{S}$ must then be exported as waste, $\mathrm W$. A balance constraint (cf. \cref{eq:balance}) at the intermediate metabolite $\mathrm{P}$ requires that:
\begin{equation}
u + v - r = 0
\label{eq:balanceP}
\end{equation}
where stoichiometric coefficients are set to 1 for simplicity. Another balance constraint at the internal energetic currency metabolite, $\mathrm{E}$, leads to:
\begin{equation}
N_\mathrm{F} u + N_\mathrm{R} r - e - y z = 0
\label{eq:balanceE}
\end{equation}
where $e$ denotes an energetic maintenance demand. The currency $\mathrm E$ is a direct precursor of biomass, at a yield $y$. Finally, the waste byproduct $\mathrm W$ is considered toxic, inducing a death rate proportional to its concentration:
\begin{equation}
\mu = z-\sigma = z-\tau w
\label{eq:death-vazquez}
\end{equation}

The parameters were set as follows. The stoichiometric coefficients $N_\mathrm{F} = 2$, $N_\mathrm{R} = 38$ are the characteristic ATP yields of glycolysis and respiration, respectively \cite{Alberts2014}. Maintenance demand is modeled as a constant drain of ATP at a rate $e = 1.0625\, \mathrm{mmol/gDW/h}$, typical of mammalian cells \cite{Kilburn1969}. The maximum respiratory capacity is computed as $r_\mathrm{max} = F_\mathrm{thr} \times \mathrm{Vol} \times \mathrm{DW} = 0.45 \, \mathrm{mmol/g/h}$, where $F_\mathrm{thr} = 0.9\, \mathrm{mM/min}$ is a glucose uptake threshold (per cytoplasmic volume) beyond which mammalian cells secrete lactate \cite{Vazquez2010b}, $\mathrm{Vol} = 3\, \mathrm{pL}$ and $\mathrm{DW} = 0.9\, \mathrm{ng}$ are the volume \cite{Zhao2008} and dry weight, respectively, of mammalian (HeLa) cells, the later estimated from the dry mass fraction ($\approx 30 \,\%$, \cite[BNID100387]{Milo2010}) and total weight ($=3\, \mathrm{ng}$ \cite{Park2008b}) of one HeLa cell. The concentration of substrate in the medium was set $c = 15\,\mathrm{mM}$, which is a typical glucose concentration in mammalian cell culture media (for example, RPMI-1640 \cite{Moore1967}). Next, $V = 0.5\,\mathrm{mmol/gDW/h}$, also measured for HeLa cells \cite{Rodriguez-Enriquez2009} (the measured flux is per protein weight, so we multiplied by 0.5 to obtain a flux per cell dry weight, since roughly half of a generic cell dry weight is protein \cite[BNID101955]{Milo2010}). The parameter $y = 348\, \mathrm{mmol/gDW}$ was adjusted so that the maximum growth rate was $\approx 1 \, \mathrm{day}^{-1}$, which is within the range of duplication rates in mammalian cells \cite{Bree1988,Dhir2000}. Finally, the toxicity of waste was set as $\tau = 0.0022\, \mathrm{h}^{-1}\mathrm{mM}^{-1}$, obtained from linearizing the death rate dependence on lactate in a mammalian cell culture reported by S. Dhir \emph{et al.} \cite{Dhir2000}.

\paragraph{Genome-scale metabolic network of CHO-K1}

Exploiting the increasingly available information about cellular metabolism at the stoichiometric level \cite{Kanehisa2014,Caspi2014}, a metabolic network can be reconstructed containing the biochemical reactions occurring inside a cell of interest. These reconstructions typically contain data about stoichiometric coefficients ($N_{ik}$, cf. \cref{eq:balance}), thermodynamic bounds ($\mathrm{lb}_k, \mathrm{ub}_k, L_i$, cf. \cref{eq:lbub,eq:u}), and a biomass synthesis pseudo-reaction ($y_i$, cf. \cref{eq:balance}) \cite{Feist2010, Schellenberger2010, Lewis2012}. Motivated by the fact that most therapeutic proteins requiring complex post-transnational modifications in the biotechnological industry are produced in Chinese hamster ovary (CHO) cell lines \cite{Wurm2004}, we analyzed the steady states of a genome scale model of the CHO-K1 line \cite{Hefzi2016}. Based on the latest consensus reconstruction of CHO metabolism available at the time of writing, containing 1766 genes and 6663 reactions, a cell-line-specific model for CHO-K1 was built by Hefzi \emph{et al.} \cite{Hefzi2016}, comprising 4723 reactions (including exchanges) and 2773 metabolites (with cellular compartmentalization). It accounts for biomass synthesis through a virtual reaction that contains the moles of each metabolite required to synthesize one gram of biomass. The network recapitulates experimental growth rates and cell-line-specific amino acid auxotrophies.

In order to enforce \cref{eq:crowd}, we complemented this network with a set of reaction costs. Following T. Shlomi \emph{et. al} \cite{Shlomi2011}, we assigned costs as follows: $\alpha_k^\pm = \mathrm{MW}_k^\pm / k_{\mathrm{cat},k}^\pm$, where $\mathrm{MW}_k^\pm$ and $k_{\mathrm{cat},k}^\pm$ are the molecular weight and catalytic rate of the enzyme catalyzing reaction $k$ in the given direction. The parameters $\mathrm{MW}_k^\pm$, $k_{\mathrm{cat},k}^\pm$ were gathered by T. Shlomi \emph{et. al} from public repositories of enzymatic data. Missing values are set to the median of available values. An estimate of the enzyme mass fraction $C = 0.078 \mathrm{mg/mgDW}$ was obtained for mammalian cells by the same authors. A constant maintenance energetic demand (cf. term $e_i$ in \cref{eq:balance}) was added in the form of an ATP hydrolysis drain at a flux rate $2.24868 \mathrm{mmol / gDW / h}$ \cite{Kilburn1969} (the reported value is for mouse LS cells, which we converted by accounting for the dry weight of CHO cells \cite{Hefzi2016}). The maximum uptake rate of glucose was set at $V_\mathrm{glc} = 0.5\,\mathrm{mmol/gDW/h}$, from previous models of cultured CHO cells fitted to experimental data \cite{Nolan2011,Kiparissides2011} (which also closely matches the values obtained from kinetic measurements on other mammalian cell lines \cite{Rodriguez-Enriquez2009}). However, kinetic parameters needed to estimate $V_i$ for most metabolites are not known at present. Based on data of CHO cell cultures in our facility (not shown), as well as data in the literature \cite{Dhir2000,Altamirano2000,Ozturk1992}, we estimated that the uptake rates of amino acids is typically one order of magnitude slower than the uptake rate of glucose, accordingly we set $V_i = V_\mathrm{glc} / 10$ for amino acids. Other metabolites have an unbounded uptake ($V_i = \infty$). In the simulations we used Iscove's modified Dulbecco's medium (IMDM), and set infinite concentrations for water, protons and oxygen.

The two toxic byproducts most commonly studied in mammalian cell cultures are ammonia and lactate. Their toxicity is primarily attributed to their effects in osmolarity and pH \cite{Bakker1996,Hu1997,Ozturk1992,Schneider1996}. It has been suggested that the accumulated toxicity may result in increased maintenance demands \cite{Lao1997,Martinelle1993} and in reduced biomass yields \cite{Martinelle1993}. Parameters describing these effects quantitatively vary over an order of magnitude \cite{Glacken1988,Gaertner1993a,Portner1996} depending on culture conditions and cell-line. In our model we incorporate these effects through the factor $K$ and for the sake of specificity in this example we use:
\begin{equation}
K = (1 + s_\mathrm{nh4} / K_\mathrm{nh4})^{-1}(1 + s_\mathrm{lac} / K_\mathrm{lac})^{-1}
\label{eq:K}
\end{equation}
with $K_\mathrm{nh4} = 1.05 \mathrm{mM}$, $K_\mathrm{lac} = 8 \mathrm{mM}$ \cite{Bree1988}, and set $\mu = K\times z$.

\subsection{Additional details}

Numerical simulations were carried out in Julia \cite{Bezanson2017}. Linear programs were solved with Gurobi \cite{gurobi}. The CHO-K1 metabolic network \cite{Hefzi2016} was read and setup with all relevant parameters using a script written in Python with the COBRApy package \cite{Ebrahim2013, Becker2007c, Schellenberger2011}. All scripts (which also include parameter values) are freely available in a public Github repository \cite{cossiogit}.

\section{Results and Discussion}

\subsection{General properties of steady states}

In this section we present the general procedure to determine the steady states of \cref{eq:dX,eq:ds} and discuss some general results of our model that are independent of the specificities of the metabolic networks of interest. The first step is to set the time-dependence in \cref{eq:dX,eq:ds} to zero,
\begin{align}
\frac{\mathrm{d} X}{\mathrm{d}t} &= (\mu - \phi D) X = 0
\label{eq:dXeq} \\
\frac{\mathrm{d} s_i}{\mathrm{d}t} &= -u_i X - (s_i - c_i)D = 0
\label{eq:dseq}
\end{align}
Note that \cref{eq:dseq} depends on $X$ and $D$ only through the ratio $1 / \xi = D / X$ (known in the literature as cell-specific perfusion rate, or CSPR \cite{Ozturk1996}), such that $\xi$ is the number of cells sustained in the culture per unit of medium supplied per unit time (the units of $\xi$ are cells $\times$ time / volume). If we recall that in our cellular model, $u_i$ is constrained by a term that also depends on $X$ and $D$ only through $\xi$ (cf. \cref{eq:u}), it immediately follows that the values of the uptakes and metabolite concentrations in steady state must be functions of $\xi$, which we denote by $u_i^*(\xi)$ and $s_i^*(\xi)$ respectively. To compute $u_i^*(\xi)$, solve the linear program of maximizing the biomass synthesis rate ($z$) subject to \cref{eq:balance,eq:lbub}, but replacing \cref{eq:u} with:
\begin{equation}
-L_i \le u_i \le \min\{V_i, c_i / \xi\}.
\label{eq:uxi}
\end{equation}
The resulting optimal value of $z$ will be denoted by $z^*(\xi)$. Moreover, once $u_i^*(\xi)$ is known, the stationary metabolite concentrations in the culture follow from \cref{eq:dseq}:
\begin{equation}
s_i^*(\xi) = c_i - u_i^*(\xi) \xi
\label{eq:seq}
\end{equation}
Then, given $z^*(\xi)$, the effective growth rate in steady state can also be given as a function of $\xi$, $\mu^*(\xi)$, by evaluating $K$ or $\sigma$ using the concentrations $s_i^*(\xi)$ from \cref{eq:seq}. Next, \cref{eq:dXeq} implies that the dilution rate at which a steady state occurs must also be a function of $\xi$, that we denote by $D^*(\xi)$. Combining this with the relation $\xi = X/D$, we obtain the steady state cell density, $X^*(\xi)$, as well:
\begin{equation}
D^*(\xi) = \mu^*(\xi) / \phi, \quad X^*(\xi) = \xi \mu^*(\xi) / \phi.
\label{eq:DXeq}
\end{equation}
Note that while \cref{eq:dXeq,eq:dseq} are satisfied by any $D \ge 0$ when $X = 0$, the value $D_\mathrm{max} = D^*(0)$ given by \cref{eq:DXeq} is actually the washout dilution rate, \emph{i.e.}, the minimum dilution rate that washes the culture of cells. Clearly all steady states with non-zero cell density are required to satisfy $D^*(\xi) < D_\mathrm{max}$.

From \cref{eq:DXeq} it is evident that the function $\mu^*(\xi)$ encodes all the information needed to get the values of $X$ in steady state at different dilution rates and for any value of the bleeding coefficient $\phi$. On the other hand, if multiple values of $\xi$ are consistent with the same dilution rate (\emph{i.e.} if the function $D^*(\xi)$ is not one-to-one), the system is multi-stable (\emph{i.e.}, multiple steady states coexist under identical external conditions). A necessary condition multi-stability is that $\mu^*(\xi)$ is not monotonously decreasing. Since the biomass production rate $z^*(\xi)$ is a non-increasing function of $\xi$ (proved in the Appendix), a change in the monotonicity of $\mu^*(\xi)$ must be a consequence of toxic byproduct accumulation, modeled through the terms $K$ and $\sigma$.

A noteworthy consequence of \cref{eq:DXeq} is that a plot displaying the parametric curve $(\phi D^*(\xi), \phi X^*(\xi))$ as a function of $\xi$ is invariant to changes in $\phi$. This means that for a given cell line and medium formulation, this curve can be obtained from measurements in a chemostat (which corresponds to $\phi = 1$), and the result will also apply to any perfusion system with an arbitrary value of $\phi$. Moreover, since $s_i^*(\xi)$ and $u_i^*(\xi)$ are independent of $\phi$, cellular metabolism in steady states is equivalent in the chemostat and any perfusion system (with an arbitrary value of $\phi$), provided that the values of $\xi = X/D$ in both systems match.

Finally, we mention that generally a threshold value $\xi_\mathrm{m}$ exists, such that a steady state with $\xi = X/D$ is feasible only if $\xi \le \xi_\mathrm{m}$. When $\xi > \xi_\mathrm{m}$, some of the constrains in \cref{eq:balance,eq:lbub,eq:u} cannot be met. In degenerate scenarios  we could have $\xi_\mathrm{m} = \infty$ (\emph{e.g.}, this could be the case if the maintenance demand in \cref{eq:balance} is neglected) or $\xi_\mathrm{m} \le 0$ (\emph{e.g.}, if the medium is so poor that the maintenance demand cannot be met even with a vanishingly small cell density). The parameter $\xi_\mathrm{m}$ arising in this way in our model, coincides with the definition of \emph{medium depth} given in the literature \cite{Konstantinov2006}, and it quantifies for a given medium composition the maximum cell density attainable per unit of medium supplied per unit time.

\paragraph{Stability}

To determine the stability of steady states we compute the Jacobian eigenvalues of the system \cref{eq:dX,eq:ds}. If the real parts are all negative the state is stable, but if at least one eigenvalue has a positive real part, the state is unstable \cite{Strogatz1994}. The critical case where all eigenvalues have non-negative real parts but at least one of them has a zero real part is dealt with using the Center Manifold Theorem \cite[§ 8.1]{Khalil2002}. The Appendix contains a detailed mathematical treatment. Briefly, an steady state is unstable if $\mu^*(\xi)$ is increasing in a neighborhood, and stable otherwise. We stated above that steady states of a given cell line in continuous culture, using a fixed medium formulation, can be given as functions of $\xi$. The condition for stability stated here is also uniquely determined by $\xi$ and, in particular, it is independent of $\phi$. Therefore, a steady state in a chemostat (with $\phi = 1$) is stable if and only if the same steady state in perfusion (with a matching value of $\xi$, but arbitrary $\phi$) is also stable. Since our results are qualitatively invariant to changes in $\phi$, we set $\phi = 1$ in what follows.

\subsection{Insight from the toy model}

\begin{figure}
	\centering
	\includegraphics[width=10cm]{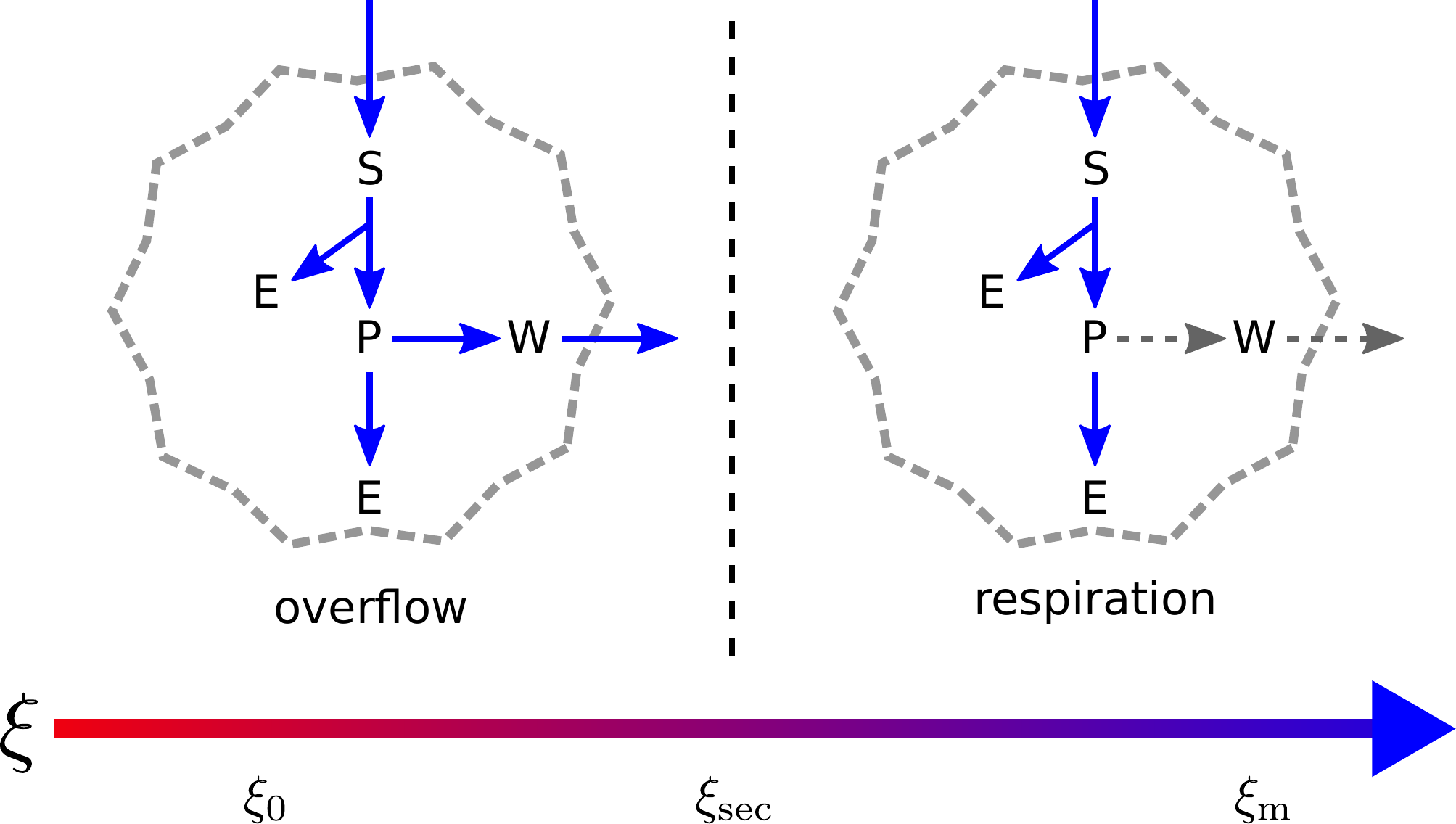}
	\caption{{\bf Metabolic modes of toy model.} The simple metabolic network depicted in \cref{fig:vazquez-net} supports two flux modes in steady state. At high enough nutrient uptake, there is an overflow regime (left) where the respiratory flux hits the upper bound $r_\mathrm{max}$ and the remaining nutrient must be exported in the form of $\mathrm W$. As a consequence less energy is produced per unit of substrate consumed. On the other hand, in the respiration regime (right) the nutrient is completely oxidized with a large energy yield. In each case, active and inactive fluxes are represented with blue arrows and discontinuous gray arrows, respectively. The bottom arrow indicates the direction of increasing $\xi$ and marks the threshold values (cf. \cref{eq:xithresh}). $\xi_0$ delimits the regimes of nutrient excess and competition. The transition between overflow metabolism and respiration occurs at the threshold $\xi_\mathrm{sec}$. Finally, maintenance demand cannot be met beyond $\xi > \xi_\mathrm{m}$. As $\xi$ grows, the biomass yield per unit of medium supplied per unit time also increases, and this is depicted by the color gradient in the arrow of values of $\xi$, going from red (low yield) to blue (high yield).}
	\label{fig:vazquez-modes}
\end{figure}

We first consider the small metabolic network depicted in \cref{fig:vazquez-net}. In this example, maximization of growth sets the nutrient uptake ($u$) and respiratory flux ($r$) at the maximum rates allowed by their respective upper bounds, \cref{eq:u,eq:rmax}. Employing \cref{eq:balanceP} to determine the waste secretion rate ($v$) from $u,r$, we obtain:
\begin{equation}
u = \min\{ V, c / \xi\}, \quad r = \min\{u, r_\mathrm{max}\}, \quad v = r - u.
\label{eq:vazquez-flux}
\end{equation}
Thus the toy model admits simple analytical expressions giving the rates of metabolic fluxes in steady states as functions of $\xi$. A minimum nutrient uptake rate $u_\mathrm{m}$ is required to sustain the maintenance demand $e$. Since most cell types are able to grow under certain conditions without waste secretion, we make the biologically reasonable assumption that $u_\mathrm{m} \le r_\mathrm{max}$ (which is satisfied by the parameters chosen in Materials and Methods). It then follows that $u_\mathrm{m} = e / (N_\mathrm{F} + N_\mathrm{R})$. There are three critical thresholds in $\xi$ that correspond to important qualitative changes in the culture:
\begin{equation}
\xi_\mathrm{m} = c / u_\mathrm{m}, \quad \xi_\mathrm{sec} = c / r_\mathrm{max}, \quad \xi_0 = c / V.
\label{eq:xithresh}
\end{equation}
These transitions can be interpreted in the following way: \textbf{1)} if $\xi \ge \xi_\mathrm{m}$ the growth rate is zero because the maintenance demand cannot be met; \textbf{2)} if $\xi_\mathrm{sec} \le \xi \le \xi_\mathrm{m}$, cells grow without secreting waste; \textbf{3)} if $\xi \le \xi_\mathrm{sec}$, there is waste secretion; \textbf{4)} for $\xi \ge \xi_0$ cells are competing for the substrate and growth is limited by nutrient availability (cf. discussion before \cref{eq:u}); \textbf{5)} finally, if $\xi \le \xi_0$ there is nutrient excess and cells are growing at the maximum rate allowed by intrinsic kinetic limitations. We emphasize that the threshold $\xi_\mathrm{sec}$ carries a special metabolic significance, because it controls the switch between two qualitatively distinct metabolic modes: if $\xi \le \xi_\mathrm{sec}$, respiration is saturated and the intermediate $\mathrm P$ \emph{overflows} in the form of secreted waste, with a lower energy yield; on the other hand, if $\xi \ge \xi_\mathrm{sec}$, the cell relies entirely on respiration to generate energy, with a higher yield (cf. \cref{fig:vazquez-modes}).

The medium carries a concentration $c$ of primary nutrient and zero waste content. Under these assumptions, \cref{eq:seq} has the following analytical solution for the steady state values of the metabolite concentrations, $s^*(\xi), w^*(\xi)$:
\begin{align}
\label{eq:vazquez-seq}
s^*(\xi) &= c - \min\{V \xi, c\} \\
\label{eq:vazquez-weq}
w^*(\xi) &= \max \left\{ 0, c - s^*(\xi) - r_\mathrm{max} \xi \right\}
\end{align}
Note that $s^*(\xi)$ is a decreasing function of $\xi$, while $w^*(\xi)$ has at most a single maximum. \cref{eq:death-vazquez,eq:vazquez-seq,eq:vazquez-weq,eq:vazquez-flux,eq:balanceE} can be used to define $\mu^*(\xi)$. Then $D^*(\xi), X^*(\xi)$ are given by \cref{eq:DXeq}.

\begin{figure}
	\centering
	\includegraphics[width=15cm]{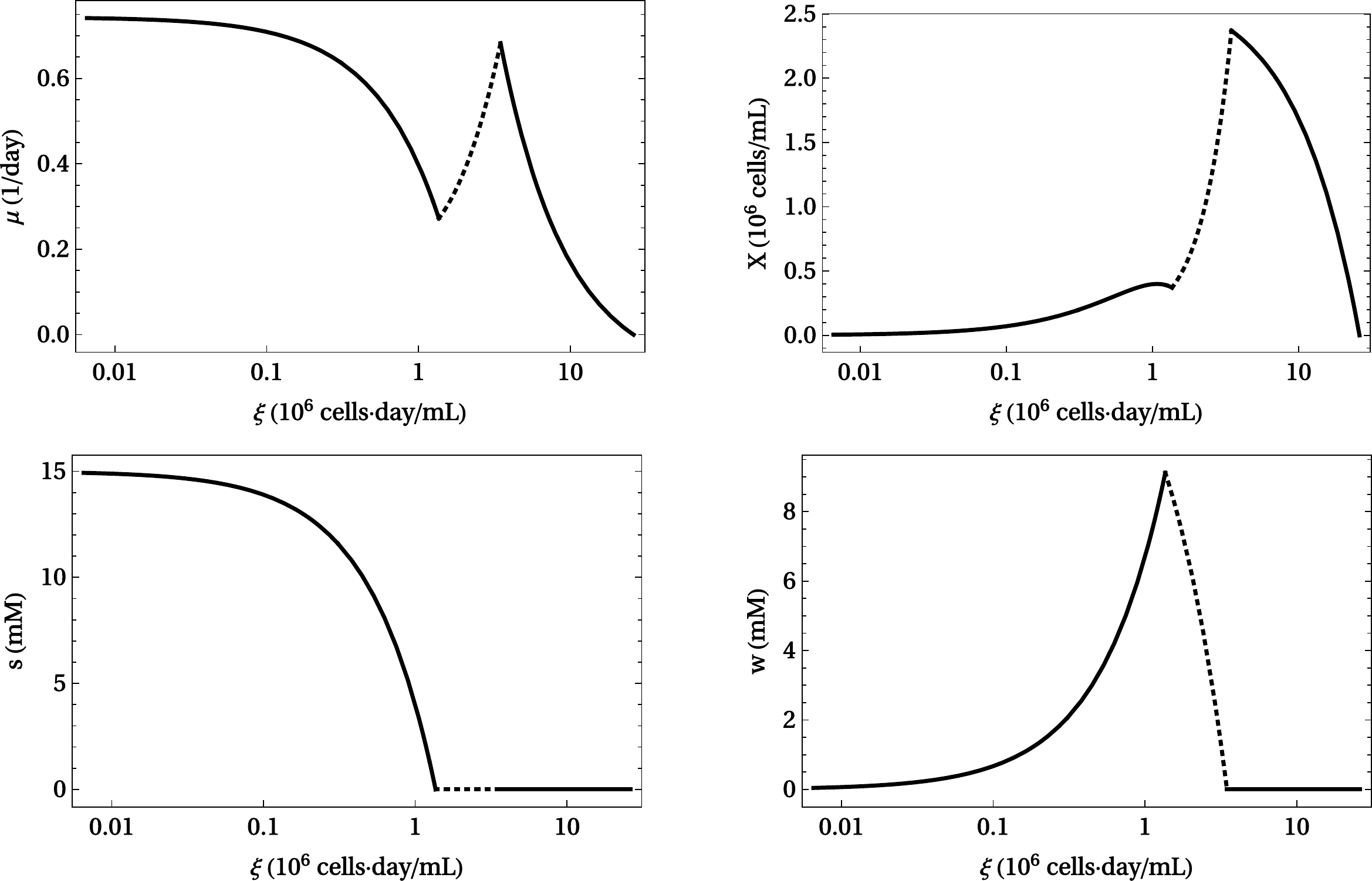}
	\caption{{\bf Steady states of toy model.} The four panels show the growth rate ($\mu$), cell density ($X$), nutrient concentration ($s$) and waste concentration ($w$) in steady states as functions of $\xi$, from $\xi = 0$ to $\xi = \xi_\mathrm{m}$. Stable states are drawn in continuous line (--), while unstable states are drawn in discontinuous line (- -). Parameter values are listed in Materials and Methods. To convert between grams of dry weight and number of cells, we used a cellular dry weight of $0.9\, \mathrm{ng/cell}$.}
	\label{fig:vazquez-eq}
\end{figure}

\cref{fig:vazquez-eq} shows plots of $\mu^*(\xi)$, $X^*(\xi)$, $s^*(\xi)$ and $w^*(\xi)$ for this  model. Parameter values are given in Materials and Methods. As $\xi$ ranges from $\xi = 0$ to $\xi = \xi_\mathrm{m}$, stable and unstable steady states are drawn in continuous and discontinuous line, respectively. The system is stable in two regimes: $\xi \lesssim 1 \times 10^6 \,\mathrm{cells} \cdot \mathrm{day/mL}$, with high toxicity, low biomass yield and low cell density; and $\xi \ge \xi_\mathrm{sec}$, with no toxicity, high biomass yield and high cell density that decays as $\xi$ increases. The later states rely solely on respiration for energy generation (\cref{fig:vazquez-modes}a), while the former states exhibit overflow metabolism (\cref{fig:vazquez-modes}b). Waste concentration initially increases with $\xi$ until a maximum value is reached. Then $w$ decays during the unstable phase, all the way to zero at $\xi_\mathrm{sec}$, where waste secretion stops and the system becomes stable again. 

Intuitively, unstable states become stressed due to high levels of toxicity, which also makes the system very sensitive to perturbations. The typical behavior of nutrients and waste products (in particular glucose and lactate, respectively) in continuous cell cultures, as observed in experiments \cite{Konstantinov2006}, is that as $\xi$ increases, nutrient concentration in the culture decreases while waste initially accumulates \cite{Konstantinov2006} but eventually phases out as cells switch towards higher-yield metabolic pathways \cite{Follstad1999, Europa2000}. This behavior is qualitatively reproduced by $s$ and $w$ in our toy model.

\begin{figure}
	\centering
	\includegraphics[width=15cm]{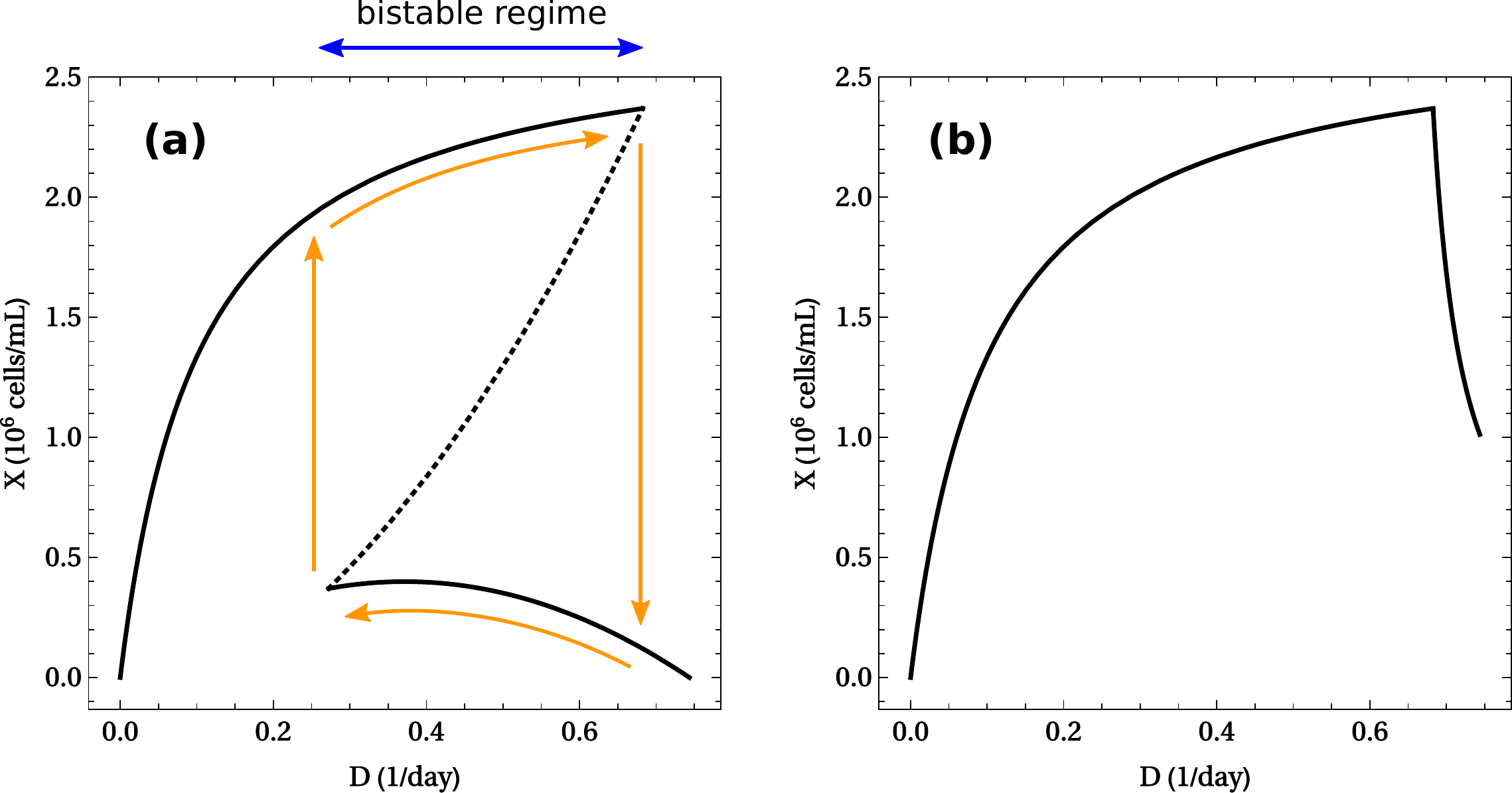}
	\caption{{\bf Cell density versus dilution rate in the solvable model.} (\textbf{a}) With toxicity ($\tau > 0$). A multi-stable regime occurs when the dilution rate is within the range $0.25 \lesssim D \lesssim 0.7$ (blue arrow). Within this range, there are three steady states compatible with the same external conditions. Two of them are stable (--) , while the other one is unstable (- -). The orange arrows trace the hysteresis loop. (\textbf{b}) Without toxicity ($\tau = 0$). The system exhibits a single stable steady state for each value of the dilution rate. Parameter values are given in Materials and Methods. To convert between grams of dry weight and number of cells, we used a cellular dry weight of $0.9\, \mathrm{ng/cell}$.}
	\label{fig:vazquez-bif}
\end{figure}

The function $\mu^*(\xi)$ is not monotonically decreasing. As explained above, this implies a coexistence of multiple steady states under identical external conditions. This is readily apparent in a bifurcation diagram of the steady values of $X$ versus $D$, as shown in \cref{fig:vazquez-bif}a. In a range of dilution rates ($0.25 \lesssim D \lesssim 0.7$, units: $\mathrm{day}^{-1}$), the system exhibits three steady states, one of which is unstable (discontinuous line in the figure), while the other two are stable (continuous line in the figure). Thus a stable state of high-cell density coexists with another of low cell density, over a range of dilution rates. Cellular metabolism in the former state is respiratory (\cref{fig:vazquez-modes}a), whereas cells in the later state exhibit an overflow metabolism (\cref{fig:vazquez-modes}b). The unstable state is an intermediate transition state lying between these two extremes.

\begin{figure}
	\centering
	\includegraphics[width=16cm]{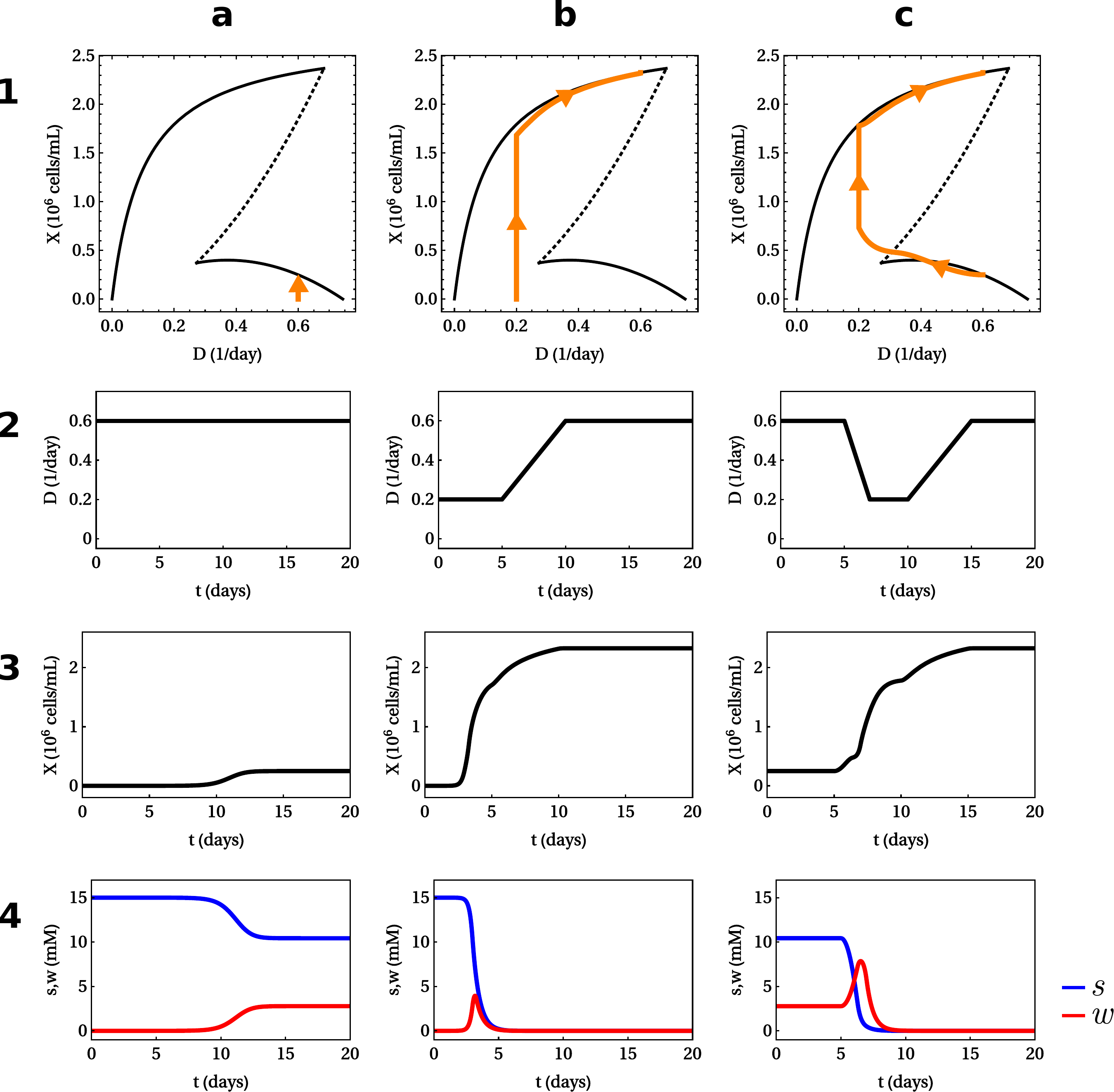}
	\caption{{\bf Dynamic simulations of toy model.} The columns (\textbf{a}, \textbf{b} and \textbf{c}) show the results of three simulations, where three different manipulations of the dilution rate in time were used, while the rows show: \textbf{1} the trajectory of the system (orange) in the $D$ vs. $X$ plane, superimposed on the bifurcation diagram (cf. \cref{fig:vazquez-bif}a); \textbf{2}, \textbf{3} and \textbf{4} the dilution rate, cell density, and metabolite concentrations in time. In all three simulations, the medium has the same nutrient concentration and the same final dilution rate, $D = 0.6 \, \mathrm{day}^{-1}$, is reached by the end of the simulation. \textbf{(a)} $D$ is constant in time, with the value $D = 0.6 \, \mathrm{day}^{-1}$. The following initial conditions are used: $X_0 = 10^{-6}$, $s_0 = c$, $w_0 = 0$. The final steady cell density is low. \textbf{(b)} $D(t)$ is a piece-wise linear function of time that interpolates between the points: $D(0) = D(5) = 0.2 \, \mathrm{day}^{-1}$, $D(10) = D(20) = 0.6 \, \mathrm{day}^{-1}$. Initial conditions are the same as in (a), but the final cell density increases five-fold. \textbf{(c)} Starting from the final steady condition in (a), the dilution rate is manipulated in time according to a piece-wise linear function that interpolates between the points: $D(0) = D(5) = 0.6 \, \mathrm{day}^{-1}$, $D(7) = D(10) = 0.2 \, \mathrm{day}^{-1}$, $D(15) = D(20) = 0.6 \, \mathrm{day}^{-1}$. The system responds by switching from the steady state in (a) to the high-cell density steady state of (b). Parameter values are listed in Materials and Methods. To convert between grams of dry weight and number of cells, we used a cellular dry weight of $0.9\, \mathrm{ng/cell}$.}
	\label{fig:vazquez-sim}
\end{figure}

Multi-stability of continuous cultures has been repeatedly observed in experiments \cite{Europa2000, Follstad1999, Altamirano2001, Hayter1992, Gambhir2003}. In our model it is a direct consequence of toxicity induced by the accumulation of waste \cite{Xiu1998}. Small variations in the  dilution rate near $D \approx 0.25 \, \mathrm{day}^{-1}$ or $D \approx 0.7 \, \mathrm{day}^{-1}$ result in discontinuous transitions where the cell density rises or drops abruptly, respectively. These jumps can be traced around an hysteresis loop, drawn in orange arrows in \cref{fig:vazquez-bif}a. More generally one may also expect that the system jumps from one state to the other due to random fluctuations. In particular, note that the basin of attraction of the high-cell density state decreases with $D$ (since the discontinuous line of unstable states eventually intersects with the high cell density states). Therefore, our toy model exhibits a plausible mechanism through which increasing dilution rates translate into high cell density states with diminishing resilience to perturbations. 

The role of toxicity becomes evident if we consider ideal cells resistant to waste accumulation (setting $\tau = 0$). The resulting plot of $X$ vs. $D$ in this case (\cref{fig:vazquez-bif}b) reveals a single stable steady state for each value of the dilution rate. There is a discontinuous transition at the washout dilution rate ($D_\mathrm{max}$), where the cell density suddenly drops to zero. Away from this value, the system is resilient to perturbations since there are no multiple steady states between which jumps can occur.

Multi-stability implies that system dynamics are non-trivial, in the sense that different trajectories might lead to different steady states. Therefore it is important for industrial applications to understand how the system is driven to one or another state. We numerically solved \cref{eq:dX,eq:ds} by performing the FBA optimization at each instant of time, in a manner analogous to DFBA \cite{Mahadevan2002}. With the parameter values given in Materials and Methods, we simulated the response of the system to three different profiles of the dilution in time. First, in \cref{fig:vazquez-sim}a, a constant dilution rate of $D = 0.6 \, \mathrm{day}^{-1}$ is used. Two possible stable steady states are consistent with this dilution rate, attaining different cell densities (cf. \cref{fig:vazquez-bif}a). Starting from a very low initial cell density, the system responds by settling at the steady state of lowest cell-density. As can be appreciated in the bottom row of \cref{fig:vazquez-sim}a, this state is characterized by an accumulation of toxic waste that prevents further cell growth. A smarter manipulation of the dilution rate makes the high-cell density state accessible. This is demonstrated in \cref{fig:vazquez-sim}b, where $D$ starts from a lower value $(D = 0.2 \, \mathrm{day}^{-1})$, and is gradually increased until the final value ($D = 0.6\, \mathrm{day}^{-1}$) is reached. The final cell density resulting from this smooth increase of the dilution rate is five-fold larger than the one obtained with a constant dilution rate. This state is also characterized by very low levels of waste accumulation (cf. last row of \cref{fig:vazquez-sim}b). We stress that external conditions in the final steady state (dilution rate and medium formulation) are the same in both cases. 

Finally, \cref{fig:vazquez-sim}c shows how the dilution rate can be manipulated to switch from one steady state to another. Starting from the final state of the simulation in \cref{fig:vazquez-sim}a, the dilution rate is first decreased to a low value ($D = 0.2\, \mathrm{day}^{-1}$), and then it is pushed back up to the starting value ($D = 0.6\, \mathrm{day}^{-1}$). The system responds by switching from the state with low cell-density to the state with high-cell density. These simulations nicely reproduce the qualitative features of the experiment performed by B. Follstad \emph{et al.}\cite{Follstad1999}, where a continuous cell culture under the same steady external conditions (dilution rate and medium) switches between different steady states by transient manipulations of the dilution rate. The response of the cell density to transient manipulations of the dilution rate best illustrated in the $X,D$ plane (cf. first row of \cref{fig:vazquez-sim}). Then it becomes obvious that the dilution rate must be pushed down to $\approx 0.2 \mathrm{day}^{-1}$, otherwise the system will not leave the low cell density state.

\subsection{Analysis of the CHO-K1 cell-line using a genome-scale metabolic network}

\begin{figure}
	\centering
	\includegraphics[width=17cm]{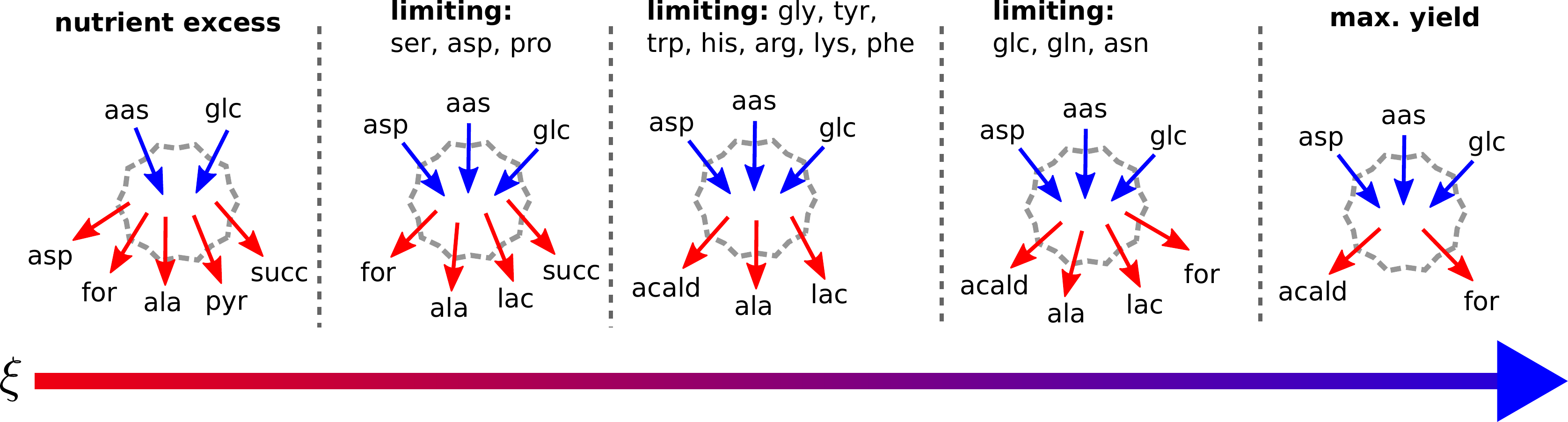}
	\caption{{\bf CHO-K1 metabolic exchange modes.} As $\xi$ increases (following the direction of the bottom arrow), CHO-K1 metabolism exhibits qualitatively distinct exchange modes. In this diagram we represent the different nutrients consumed by the cell (blue arrows) and byproducts secreted (red arrows) at different stages, while at the top we annotate the nutrients that become limiting. As $\xi$ grows, the biomass yield per unit of medium supplied per unit time also increases, and this is depicted by the color gradient in the arrow of values of $\xi$, going from red (low yield) to blue (high yield).}
	\label{fig:cho-modes}
\end{figure}

\begin{figure}
	\centering
	\includegraphics[width=17cm]{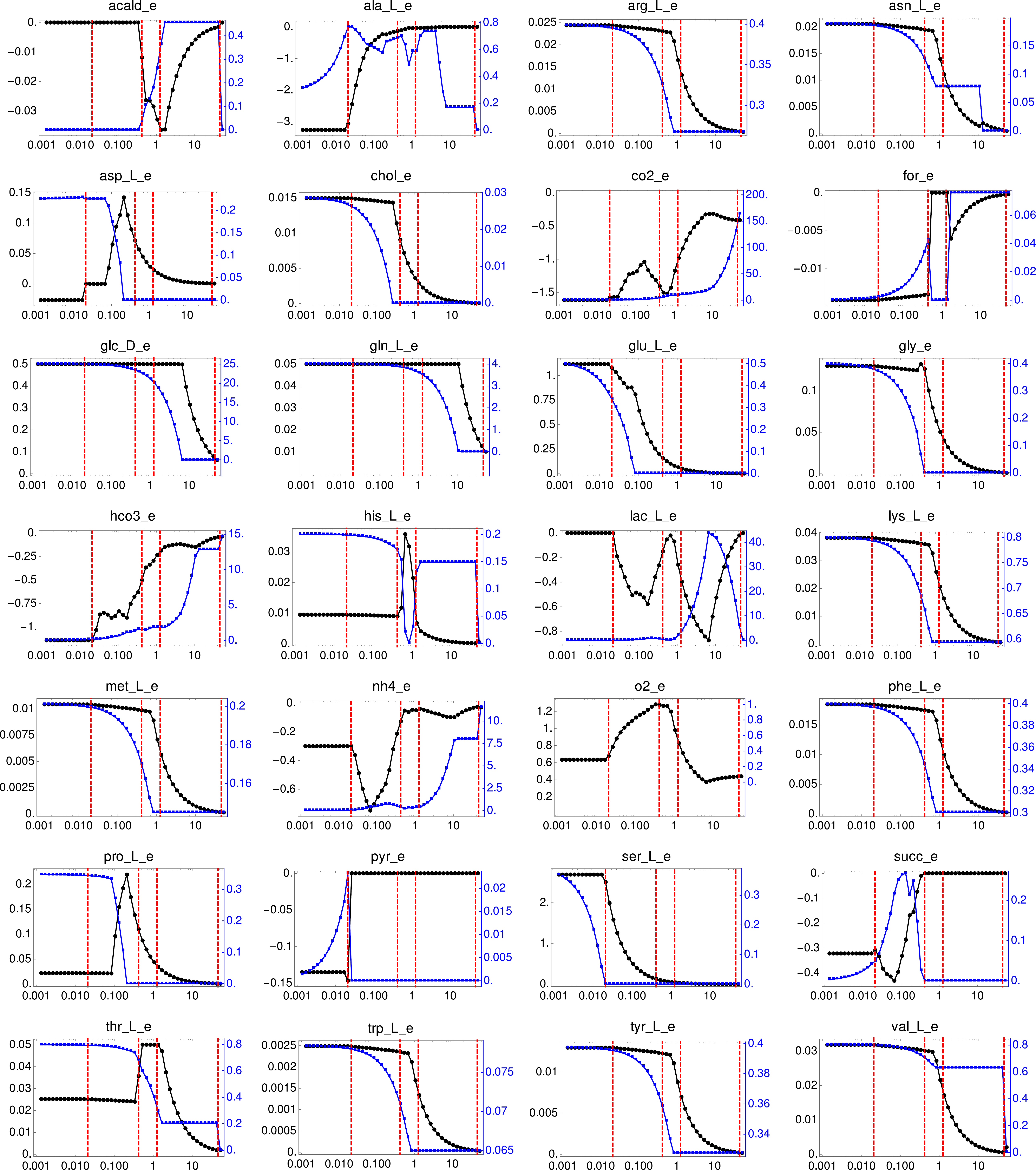}
	\caption{{\bf Uptake rates and concentrations of selected extracellular metabolites as functions of $\xi$.} Using the metabolic model of CHO-K1 and the media IMDM, we plot metabolite uptakes (black, left-axis; units: mmol/gDW/hr) and culture concentrations (blue, right-axis; units: mM) as functions of $\xi$ (units: cells $\cdot$ day / mL) for selected metabolites. The red lines indicate the transitions depicted in \cref{fig:cho-modes}. Metabolite names are the same as in the metabolic network \cite{Hefzi2016}. Abbreviations: acetaldehyde (\emph{acald\_e}), alanine (\emph{ala\_L\_e}), asparagine (\emph{asn\_L\_e}), aspartate (\emph{asp\_L\_e}), choline (\emph{chol\_e}), formate (\emph{for\_e}), glucose (\emph{glc\_D\_e}), glutamine (\emph{gln\_L\_e}), glutamate (\emph{glu\_L\_e}), glycine (\emph{gly\_e}), histidine (\emph{his\_L\_e}), lactate (\emph{lac\_L\_e}), methionine (\emph{met\_L\_e}), phenylalanine (\emph{phe\_L\_e}), proline (\emph{pro\_L\_e}), pyruvate (\emph{pyr\_e}), serine (\emph{ser\_L\_e}), succinate (\emph{succ\_e}), threonine (\emph{thr\_L\_e}), valine (\emph{val\_L\_e}).}
	\label{fig:cho-us}
\end{figure}

We determined the steady states of a continuous cell culture of the CHO-K1 line. Cellular metabolism was modeled using the reconstruction given by Hefzi \emph{et al.} \cite{Hefzi2016}, the most complete available at the time of writing. In the simulations we used Iscove's modified Dulbecco's growth media (IMDM), which is typically employed in mammalian systems. Similar to what we found in the toy model (cf. \cref{fig:vazquez-modes}), and in qualitative agreement with experimental observations \cite{Gambhir2003, Europa2000}, cells exhibited several metabolic transitions between distinct flux modes as $\xi$ was varied. However, in contrast to the the toy model, the CHO-K1 genome-scale metabolic network displays a rich multitude of transitions, as expected from its greater complexity. Because of their importance in the performance of the culture, we focused on metabolic changes that have an impact on macroscopic properties of the bioreactor, \emph{i.e.}, those that affect metabolite exchanges with the extracellular media ($u_i$). 

Although many classifications are possible, we organized our discussion by focusing on five qualitatively different modes based on the secretion of lactate and formate. \cref{fig:cho-modes} shows cartoon diagrams of these phases in order of increasing $\xi$. On the top of each diagram we annotate the nutrients that became limiting for growth during a phase. Blue arrows indicate consumption and red arrows secretion. We focused on metabolites that changed their role between phases. In particular, $\mathrm{NH}_4$ was secreted in all phases and therefore was omitted from the figure to reduce clutter. A more detailed representation of our results is given in \cref{fig:cho-us}, which shows metabolite concentrations ($s_i$) and uptakes ($u_i$) in steady states as functions of $\xi$ for a sub-set of selected metabolites. Red lines in these plots indicate the transitions depicted in \cref{fig:cho-modes}.

For small values of $\xi$ we found that glucose and almost all the amino acids available in the media were consumed, but none of them reached limiting concentrations. We call this the \emph{nutrient excess} phase, where substrate uptake is limited only by intrinsic kinetic properties of cellular transporters. Remarkably lactate was not secreted at this stage, since pyruvate was converted instead to alanine \cite{Ozturk1992} (although a small fraction of pyruvate was secreted as well \cite{ODonnell-Tormey1987}). The cell also produced succinate \cite{Duarte2014} and formate, the later being an overflow product of one-carbon metabolism of serine and glycine \cite{Meiser2016b}.

As $\xi$ continues to increase, the first metabolite that becomes limiting is serine. This marks the end of the nutrient excess phase, coinciding also with the onset of lactate secretion. At this point pyruvate is no longer secreted into the culture. Remarkably, aspartate switches from being a secreted byproduct in the first phase \cite{Duarte2014}, to consumption. Even more striking is that the specific uptake rate of aspartate and proline quickly increase until both reach limiting concentrations. A third phase is entered when succinate and formate production ceases, coinciding with a limitation of glycine. Histidine consumption rises steeply until it too reaches limiting concentrations. Other nutrients that limit growth include tyrosine, tryptophan, arginine, lysine and phenylalanine. This phase is also characterized by secretion of acetaldehyde. Remarkably, formate secretion is resumed in a later phase, where glucose, glutamine and asparagine also become limiting.

Finally, as $\xi$ approaches the maximum value $\xi_\mathrm{m}$, lactate and alanine secretion cease. This ideal state attains the highest possible biomass yield per unit of medium supplied per unit time. Note that the increase of $\xi$ has brought an overall qualitative switch to a state of metabolic efficiency where the number of secreted byproducts has dropped significantly, compared to the nutrient excess phase. Notably, the cell-specific ammonia secretion was sustained even in the states of highest biomass yield, indicating a nitrogen imbalance. This behavior has been seen qualitatively in some experiments. For example, using a CHO-derived cell line \cite{Altamirano2001}, secretion of ammonia was sustained even after a transition to an efficient metabolic phenotype with low lactate secretion and high cellular yields. However this observation seems to be cell-line dependent, and in another experiment with an hybridoma, ammonia accumulation decreased with increasing $\xi$ \cite{Konstantinov2006}.

All of the secreted metabolites predicted by our model have been verified in experiments in mammalian cell cultures \cite{ODonnell-Tormey1987, Ozturk1992, Meiser2016b, Duarte2014}, with the exception of acetaldehyde. For this metabolite our search in the literature did not reveal any experimental evidence refuting or validating the presence of this byproduct in mammalian cell culture.

\begin{figure}
	\centering
	\includegraphics[width=16cm]{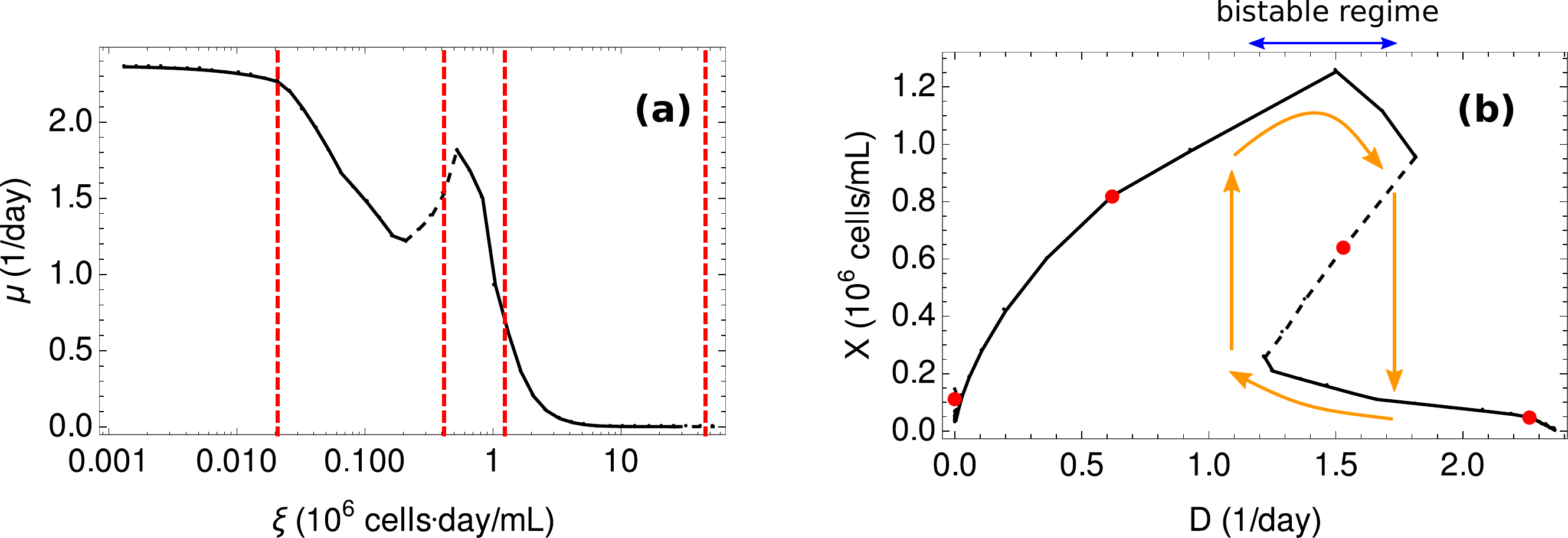}
	\caption{{\bf Steady states of CHO-K1.} {\bf (a)} Growth rate in steady states as a function of $\xi$. {\bf (b)} Bifurcation diagram of cell density versus growth rate. A bistable regime emerges as a consequence of toxic byproduct accumulation. The hysteresis loop is highlighted with orange arrows. The metabolic transitions depicted in \cref{fig:cho-modes} are depicted with red lines (in (a)) and red dots (in (b)).}
	\label{fig:cho-xi}
\end{figure}

The performance of cell-lines and media are typically evaluated by measurements performed in batch experiments \cite{Reinhart2015}. Measurements performed in the exponential phase of batch only reveal the behavior of continuous cultures at very low $\xi$, in conditions of nutrient excess. The existence of a rich multitude of qualitatively distinct metabolic behaviors at higher values of $\xi$ is missed by these experiments and therefore the assessment should not be extrapolated to high-density perfusion systems. As our analysis reveals, several nutrients may switch from basal rates of consumption to growth limiting at later values of $\xi$, while others go from secreted byproducts to consumption \cite{Martinez2013}. These examples indicate that nutrients could be in excess in a batch experiment but need not be so in the ideal regime of high-cell density perfusion cultures, at high $\xi$. Our model suggests that a better characterization of a cell-line and media formulation can be obtained in a chemostat, since the full spectrum of values of $\xi$ can be explored in this device and it faithfully reproduces all the metabolic transitions found in perfusion.

The effects of toxic byproduct accumulation are explored in \cref{fig:cho-xi}. Although the model can easily accomodate any number of toxic compounds, we considered 

We considered the toxic effects of the most commonly studied metabolites in this regard: lactate and ammonia, although the model can easily accommodate the effects of additional toxic compounds if necessary. In \cref{fig:cho-xi}a we plot the effective growth rate, $\mu$ as function of $\xi$. Stable states are drawn in continuous line, unstable states are dashed and the red dots indicate the metabolic transitions depicted in \cref{fig:cho-modes}. Note that $\mu^*(\xi)$ is not monotonous. In particular, metabolic transitions resulting in lactate and ammonia secretion peaks produce a sink in the curve $\mu^*(\xi)$. On the other hand, metabolic transitions associated to the secretion of other non-toxic byproducts do not imply changes in the monotonicity of $\mu^*(\xi)$.

The non-monotonicity of $\mu^*(\xi)$ results in multiple stable states coexisting at the same dilution rate, as evident in the bifurcation diagram \cref{fig:cho-xi}b. This resonates with the results obtained in the simpler model considered above, and is also consistent with many experimental observations of bi-stability in the literature \cite{Europa2000, Follstad1999, Altamirano2001, Hayter1992, Gambhir2003}. The regime with high-cell density corresponds to a higher value of $\xi$ and exhibits a lower accumulation of toxic byproducts (lactate and ammonia). Metabolism in this regime is also more efficient, with less byproducts secreted (cf. \cref{fig:cho-modes}). On the other hand, low cell density states are wasteful, with high levels of environmental toxicity preventing further cell growth. Again, bi-stability implies the existence of an hysteresis loop (orange arrows in the figure), where the system may suffer abrupt transitions between high and low cell densities.

\section{Concluding remarks}

In this work we have presented a model of cellular metabolism in continuous cell culture. Although similar in spirit to DFBA, our dynamic equations include terms accounting for the continuous medium exchange that enables steady states in this system. We presented a simple method to compute the steady states of the culture as a function of the ratio between cell density and dilution rate ($\xi = X / D$), scalable to metabolic networks of arbitrary complexity. In the literature $1/\xi$ is known as the cell-specific perfusion rate (CSPR), introduced by S. Ozturk \cite{Ozturk1996} who already made the empirical observation that control of the CSPR can be used to maintain a constant cell environment independent of cell growth \cite{Ozturk1996}. Our model theoretically supports this idea and leads to a stronger conclusion: that for a given cell line and medium formulation, the steady state values of the macroscopic variables of the bioreactor are all unequivocally determined by $\xi$. Therefore, $\xi$ is an ideal control parameter to fix a desired steady state in a continuous cell culture.

The model is consistent with \emph{multi-stability}, a phenomenon repeatedly observed in experiments in continuous cell cultures where multiple steady states coexist under identical external conditions. Moreover, our model accounts for metabolic switches between flux modes, experimentally observed in continuous cell culture in response to variations in the dilution rate \cite{Hayter1993}. These transitions affect the consumption or secretion of metabolites and the set of nutrients limiting growth. As a consequence, the metabolic landscape of steady states in perfusion cell cultures is complex and cannot be reproduced in batch cultures. This has the practical implication that assesments of medium quality and cell line performance carried out in batch \cite{Reinhart2015} should not be extrapolated to perfusion, since they might be missleading in this setting.

However, our analysis reveals a simple scaling law between steady states in the chemostat and any perfusion system. The landscape of metabolic transitions in the later system can be faithfully reproduced in the chemostat. Thus, for a fixed cell-line and medium formulation, the diagram displaying the values of $\phi X$ versus $\phi D$ in steady state is invariant across perfusion systems with any bleeding ratio ($\phi$), cf. \cref{eq:DXeq}, while metabolism is equivalent if the ratio $\xi = X / D$ is the same. The practical consequence is that the chemostat is an ideal experimental model where cell-lines and medium formulations can be benchmarked for their performance in high-cell density industrial continuous cultures.

Further, the model predicts that multi-stability is a consequence of negative feedback on cell growth due to accumulation of toxic byproducts in the culture. The qualitative complexity of the $\phi X$ versus $\phi D$ diagram depends only on the behavior toxic metabolites. Moreover, multi-stability implies that the system is sensitive to initial conditions and transient manipulations of external parameters. In practice, the dilution rate must be manipulated carefully to bring the system to a desired state. Thus, starting from a seed of low cell density, sharp increases of the dilution may land the system on a steady state of high toxicity and low biomass. On the other hand, slowly increasing the dilution rate will surely lead towards high-cell density states.

The conclusions stated above rely on the validity of our assumptions. In particular, we have considered a homogeneous cell population in a well-mixed bioreactor. Both assumptions are behind many models published in the field and provide reasonable fits to experimental data \cite{BenYahia2015}. Mechanical stirring of the culture typically achieves a well-mixed solution, but care must be taken to prevent mechanical damage to the cells \cite{Hu2011} (but see Ref \cite{Nienow2006}). Moreover, that the cell population can be treated attending only to its average properties is justified by the large number of cells in a typical culture ($\sim 10^6$ -- $10^8$ cells / mL), although in some settings cell-to-cell heterogeneity might become relevant \cite{Delvigne2014}. Next, to develop a specific model of cellular metabolism, we adopted a flux-balance approach \cite{Edwards2002}, where cells are assumed to optimize their metabolism towards growth rate maximization. Although this framework is well supported in the literature \cite{Palsson2006}, it is worth noting that we did not consider the kinetics of intracellular metabolites or additional regulatory mechanisms that may also control metabolic fluxes. Additionally, the quantitative predictions of the model rely on the accuracy of parameters found in the literature and databases. Among these, the flux cost coefficients ($\alpha_i$, \cref{eq:crowd}) are not available for many enzymes. If too many of these parameters are absent, calculations from FBA might be degenerate \cite{Mahadevan2003, Noor2016a}. Another important omission from the present model is that we did not consider explicitly the exchange between the culture and a gaseous phase. In particular, this includes oxygen exchange. Therefore our approach is only valid if this exchange does not become limiting to cellular growth. Despite these limitations, we have shown that the model predictions are in qualitative agreement with experimental data. More importantly, the conclusions stated above are \emph{independent} of the values of model parameters.

\section{Acknowledgements}

This project has received funding from the European Union's Horizon 2020 research and innovation programme MSCA-RISE-2016 under grant agreement No. 734439 INFERNET. The authors warmly thank Tamy Boggiano and Ernesto Chico for many helpful discussions and for reviewing this manuscript.

\section{Author contributions}

All authors contributed equally to model design; all authors wrote the manuscript; J.F.C.D. performed simulations; all authors analyzed results.

\section{Competing financial interests}

The authors declare no competing financial interests.

\bibliography{ref-fix.bib}

\clearpage

\appendix
\section{Supplementary Materials}

\subsection{Stability of fixed points}

The growth rate $\mu$ depends on the biomass synthesis rate, $z$, and on the concentrations of toxic metabolites in the culture. Since for a fixed dilution rate, $z$ depends only on $X$ (cf. \cref{eq:u}), it follows that we can write the growth rate as a function of $X$ and $\underline s$, thus $\mu = \mu(X, \underline s)$.

In particular note that the dynamics of non-toxic metabolites can be decoupled from the rest of the system (cf. \cref{eq:dX,eq:ds}). It is enough to determine the stability of a reduced system, where only $X$ and the concentrations $s_i$ of metabolites $i$ that are toxic intervene. In the trivial case where there are no toxic metabolites all fixed points are stable because $\mu$ is a non-increasing function of $X$. We assume that $\partial \mu / \partial s_i < 0$ for all toxic metabolites $i$.

Let us begin by defining the velocities of change of $X$ and $s_i$ as the right-hand sides of \cref{eq:dX} and \cref{eq:ds}, respectively,
\begin{align}
\label{eq:F}
F(X, \underline s) & = (\mu - \phi D) X \\
\label{eq:Gi}
G_i(X, \underline s) & = -u_i X - (s_i - c_i) D
\end{align}
A fixed point $\hat X, \hat{\underline s}$ satisfies $F(\hat X, \hat{\underline s}) = 0$ and $G_i(\hat X, \hat{\underline s}) = 0$. To determine its stability, we evaluate the Jacobian ($\mathbf{J}$) of \cref{eq:F,eq:Gi} at $\hat X, \hat{\underline s}$:
\begin{equation}
\mathbf{J}=\left(\begin{array}{cc}
\frac{\partial F}{\partial X} & \frac{\partial F}{\partial s_i}\\
\frac{\partial G_j}{\partial X} & \frac{\partial G_j}{\partial s_i}
\end{array}\right)=\left(\begin{array}{ccccc}
\frac{\partial\mu}{\partial X}\hat{X} & -\frac{\partial\mu}{\partial s_1}\hat{X} & -\frac{\partial\mu}{\partial s_2}\hat{X} & \cdots & -\frac{\partial\mu}{\partial s_m}\hat{X}\\
-u_{1}\left(\hat{X}\right)-u_{1}^{'}\left(\hat{X}\right)\hat{X} & -D & 0 & \cdots & 0\\
-u_{2}\left(\hat{X}\right)-u_{2}^{'}\left(\hat{X}\right)\hat{X} & 0 & -D & \cdots & 0\\
\vdots & \vdots & \vdots & \ddots & \vdots\\
-u_{m}\left(\hat{X}\right)-u_{m}^{'}\left(\hat{X}\right)\hat{X} & 0 & 0 & \cdots & -D
\end{array}\right)
\label{eq:J}
\end{equation}
where $u_i'(\hat X)$ are the derivatives of $u_i(X)$ with respect to $X$, evaluated at the fixed point. To evaluate $\partial \mu / \partial X$, recall that $\mu = K\times z - \sigma$, where $K$ and $\sigma$ depend only on $\underline s$, while $z$ is a function of $X$ (at a fixed dilution rate, cf. \cref{eq:u}). Therefore, it is enough to evaluate $z'(X)$.

\paragraph{Computation of $z'(\hat X)$ and $u_i'(\hat X)$}
Before continuing, let us make a short digression into Linear Programming \cite{Vanderbei2014}. To determine a \emph{basic} solution of FBA, it is enough to specify: (i) the indexes of fluxes that are away from their lower and upper bounds, and (ii) for the remaining fluxes, whether they are equal to their lower or upper bound. This information is called the \emph{basis} \cite{Vanderbei2014}. The full solution can be reconstructed from knowledge of the \emph{basis} by solving the linear equality constrains. Since the basis is a discrete object, it will remain constant as $\xi$ varies continuously, except for discrete `critical' values of $\xi$ where the basis changes. When the basis remains constant, $u_i^*(\xi)$ and $z^*(\xi)$ have the following forms:
\begin{equation}
u_i^*(\xi) = \alpha_i + \beta_i / \xi, \quad z^*(\xi) = \alpha + \beta / \xi,
\label{eq:ansatz}
\end{equation}
where $\alpha_i,\beta_i,\alpha,\beta$ are constant as long as the basis remains fixed. \cref{eq:ansatz} is simply the generic affine dependency on the upper bounds of the uptakes (cf. \cref{eq:uxi}). Since $z^*(\xi)$ is a non-increasing function of $\xi$, $\beta \ge 0$. Using \cref{eq:ansatz}, it follows that $u_i(\hat X) = \alpha_i + \beta_i D / \hat X$ and $z(\hat X) = \alpha + \beta D / \hat X$. Therefore:
\begin{equation}
\hat u_i + u_i'(\hat X) \hat X = \alpha_i, \quad z'(\hat X) \hat X = -\beta / \xi.
\label{eq:shadow}
\end{equation}
To obtain $\alpha,\beta,\alpha_i,\beta_i$, we exploit the fact that we will be computing $\mu^*(\xi) = z(X^*(\xi))$, $u_i^*(\xi) = u_i(X^*(\xi))$ and $X^*(\xi)$ over a sequence of contiguous values of $\xi$. If $\xi_1, \xi_2$ are sufficiently nearby:
\begin{equation}
\begin{alignedat}{4}
&\alpha_i &&= \frac{u_i^*(\xi_1) \xi_1 - u_i^*(\xi_2) \xi_2}{\xi_1 - \xi_2}, \quad
&\beta_i  &&= -\frac{u_i^*(\xi_1) - u_i(\xi_2)}{\xi_1 - \xi_2}  \xi_1 \xi_2, \\
&\alpha   &&= \frac{z^*(\xi_1) \xi_1 - z^*(\xi_2) \xi_2}{\xi_1 - \xi_2}, \quad
&\beta    &&= -\frac{z^*(\xi_1) - z^*(\xi_2)}{\xi_1 - \xi_2}  \xi_1 \xi_2.
\end{alignedat}
\label{eq:ab}
\end{equation}
The singular case $\hat X = 0$ has $\beta = \beta_i = 0$, assuming that for very low cell densities growth is not limited by substrate availability (\emph{i.e.}, that the medium is rich; cf. discussion before \cref{eq:u}).

From $z'(\hat X)$ we compute $\partial \mu / \partial X = K \times z'(\hat X)$.

\paragraph{Stability of the linearized system}
The system is stable if the real parts of all the eigenvalues of $\mathbf J$ are negative, and is unstable if at least one eigenvalue has a positive real part \cite{Strogatz1994}. The eigenvalues of $\mathbf J$ are:
\begin{equation}
\lambda_\pm = \frac{1}{2} \left( \mu'(\hat X) \hat X - D \pm 
\sqrt{(D + \mu'(\hat X) \hat X)^2 + 4\hat X \omega} \right),
\quad \lambda_\mathrm{d} = -D
\label{eq:eigvals}
\end{equation}
where $\mu'(\hat X)$ denotes the derivative $\partial \mu / \partial X$ evaluated at $\hat X$, and
\begin{equation}
\omega = -\sum_i \frac{\partial \mu}{\partial s_i}
\left( \hat u_i + u_i'(\hat X) \hat X \right) = 
- \sum_i \frac{\partial \mu}{\partial s_i} \alpha_i.
\label{eq:gamma}
\end{equation}
$\lambda_\mathrm{d}$ is a degenerate eigenvalue of order $m - 1$ (where $m$ is the number of metabolites) and is always negative (we assume that $D > 0$). The couple $\lambda_\pm$ forms a complex conjugate pair with negative real part if $(D + \mu'(\hat X) \hat X)^2 + 4\hat X \omega < 0$, which implies $\omega < 0 < \hat X$. In this case the system is stable. If $(D + \mu'(\hat X) \hat X)^2 + 4\hat X \omega \ge 0$ all the eigenvalues are real and all are negative except possibly $\lambda_+$. After some algebra, we find that $\lambda_+ < 0$ (the system is stable) or $\lambda_+ > 0$ (the system is unstable) according to whether $-\mu'(\hat X) \hat X > \xi \omega$ or $-\mu'(\hat X) \hat X < \xi \omega$, respectively. Since $\omega < 0 < \hat X$ implies $-\mu'(\hat X) \hat X > \xi \omega$ (because $\mu'(\hat X) \le 0$), the condition $-\mu'(\hat X) \hat X > \xi \omega$ is sufficient for stability, while $-\mu'(\hat X) \hat X < \xi \omega$ is sufficient for instability, even if $\lambda_\pm$ turn out to be complex.

The critical case $\lambda_+ = 0$ occurs whenever $-\mu'(\hat X) \hat X = \xi \omega$. In this case the stability of the system cannot be resolved by analysis of the linearized system alone, and we must recur to the Center Manifold Theorem \cite[Sec. 8.1]{Khalil2002}. As will be shown below, in this case the system is stable. Therefore, the fixed point is stable if $-\mu'(\hat X) \hat X \ge \xi \omega$ and unstable if $-\mu'(\hat X) \hat X < \xi \omega$. Since $\mu'(\hat X) \hat X$ and $\omega$ are both independent of $\phi$ (by \cref{eq:shadow}), this condition does not depend on $\phi$. Then, whether a fixed point is stable or not can be given as a function of $\xi$ only, as asserted in the main text.

The condition for stability can be further simplified by noting that $\mu'(\hat X) \hat X / \xi + \omega$ is the derivative of $\mu^*(\xi)$ with respect to $\xi$. Therefore, the system is stable if an only if $\mu^*(\xi)$ is non-increasing in a neighborhood.

\paragraph{Center manifold stability for the critical case ($\lambda_+ = 0$)}

If $-\mu'(\hat X) \hat X = \xi \omega$ all eigenvalues are real and negative except $\lambda_+ = 0$. In this case the linearized system cannot be used to determine the stability of the fixed point, because the effect of small perturbations along the direction of the eigenvector corresponding to $\lambda_+$ (the so-called center manifold) is not captured by the linearized system. Since only one eigenvalue has a zero real part, the Center Manifold Theorem \cite[Sec. 8.1]{Khalil2002} can be used to find a reduced one-dimensional system where the stability can be determined. For simplicity we will only consider the case where $\mu'(\hat X) = \alpha_i = 0$. The eigenvectors of $\mathbf J$ then are:
\begin{equation}
\underline p_1 = \left[\begin{array}{c}
1 \\ 0 \\ \vdots \\ 0
\end{array}\right], \quad
\underline p_2 = \left[\begin{array}{c}
-\xi \frac{\partial\mu}{\partial s_1} \\
1 \\ 0 \\ \vdots \\ 0
\end{array}\right], \quad
\underline p_3 = \left[\begin{array}{c}
-\xi \frac{\partial \mu}{\partial s_2} \\
0 \\ 1 \\ \vdots \\ 0
\end{array}\right], \quad \dots,
\underline p_{m+1} = \left[\begin{array}{c}
-\xi\frac{\partial\mu}{\partial s_{m}} \\
0 \\ 0 \\ \vdots \\ 1
\end{array}\right]
\end{equation}
where $\underline p_1$ corresponds to $\lambda_+$, $\underline p_2$ to $\lambda_-$, and the rest to $\lambda_\mathrm{d}$. These eigenvectors are assembled into a similarity matrix $\mathbf M$ (as columns), which serves to diagonalize $\mathbf J$:
\begin{equation}
\mathbf{M}^{-1}\mathbf{J}\mathbf{M}=\left[\begin{array}{ccccc}
\lambda_{+} & 0 & 0 & \cdots & 0\\
0 & \lambda_{-} & 0 & \cdots & 0\\
0 & 0 & \lambda_{\mathrm{d}} & \cdots & 0\\
\vdots & \vdots & \vdots & \ddots & \vdots\\
0 & 0 & 0 & \cdots & \lambda_{\mathrm{d}}
\end{array}\right]
\end{equation}
Introduce new variables $x,z_1,\dots,z_m$ through the relation:
\begin{equation}
\left[\begin{array}{c}
X \\ \underline s\\
\end{array}\right] = \left[\begin{array}{c}
\hat X \\ \hat {\underline s} 
\end{array}\right] + \mathbf M \left[\begin{array}{c}
x \\ \underline z
\end{array}\right]
\end{equation}
To find the reduced system, we set $\underline z = 0$, which implies $X = \hat X + x$ and $\underline s = \underline{\hat s}$. Then, differentiating $x$ with respect to time:
\begin{equation}
\frac{\mathrm d x}{\mathrm d t} = (\mu - \phi D)(\hat X + x)
\label{eq:red}
\end{equation}
The system is stable if and only if \cref{eq:red} is stable at $x = 0$. We show now that the right-hand side of \cref{eq:red} is a decreasing function of $x$, which implies stability. Since $s_i = \hat s_i$ is fixed, $K,\sigma$ are constant. From \cref{eq:ansatz} we know that $z = \alpha + \beta D / (\hat X + x)$ with constant $\alpha,\beta$ for sufficiently small $x$. Since $\hat X$ is a fixed point, it follows that $K(\alpha + \beta D / \hat X) = \sigma + \phi D$. Therefore the right-hand side of \cref{eq:red} is:
\begin{equation}
K\left(\alpha + \beta \frac{D}{\hat X + x} - \alpha - \beta \frac{D}{\hat X}\right) (\hat X + x)
= - K \beta x / \xi
\end{equation}
which is decreasing in $x$. This argument breaks down if $\beta = 0$, which occurs only in conditions of nutrient excess, where $\xi$ is low enough that there is no nutrient competition between the cells. In this case $\beta_i = 0$ also for all $i$, implying that $\hat u_i = \alpha_i$ is piece-wise constant in this regime. If toxic metabolites are being secreted, $\omega > 0$ implying $-\mu'(\hat X) \hat X = 0 < \xi \omega$, which falls under the umbrella of the non-critical linear stability analysis discussed above. If toxic metabolites are not being secreted, $\omega = 0$. But in the later case $X$ is uncoupled from the rest of the variables, and the system is trivially stable because $\mu = \alpha$ is constant.

\end{document}